\documentclass[12pt,amsmath,twocolumn]{aastex631}
\usepackage{graphicx}
\usepackage{bm}
\usepackage{epsfig}
\usepackage{color}
\usepackage{ulem}
\usepackage{hyperref}
\usepackage{cancel}
\usepackage{soul}
\usepackage[mathscr]{euscript}

\hypersetup{
    colorlinks=true,
    linkcolor=red,
    citecolor=blue,
} 

\input{epsf}
\newcommand{\beq}{\begin{equation}}
\newcommand{\eeq}{\end{equation}}
\newcommand{\bea}{\begin{eqnarray}}
\newcommand{\eea}{\end{eqnarray}}
\newcommand{\bi}{\begin{itemize}}
\newcommand{\ei}{\end{itemize}}
\newcommand{\bfi}{\begin{figure}[!t]
\epsfxsize=7cm
\epsffile}
\newcommand{\bfib}{\begin{figure}[htb]
\epsfxsize=9cm
\epsffile}
\newcommand{\bfig}{\begin{figure*}[htb]
\epsfxsize=12cm
\epsffile}
\newcommand{\efi}{\end{figure}}
\newcommand{\efib}{\end{figure}}
\newcommand{\efig}{\end{figure*}}
\newcommand{\no}{\nonumber}
\newcommand{\bfs}{\mbox{\boldmath$s$}}
\newcommand{\bfr}{\mbox{\boldmath$r$}}
\newcommand{\bfx}{\mbox{\boldmath$x$}}
\newcommand{\bfk}{\mbox{\boldmath$k$}}
\newcommand{\bfv}{\mbox{\boldmath$v$}}
\newcommand{\bfu}{\mbox{\boldmath$u$}}

\newcommand{\bfq}{\mbox{\boldmath$q$}}

\newcommand{\HI}{\mathrm{HI}}
\newcommand{\Htr}{\mathrm{H}}

\newcommand{\Pdd}{P_{\delta\delta}}
\newcommand{\Pdt}{P_{\delta \Theta}}
\newcommand{\Ptt}{P_{\Theta\Theta}}

\newcommand{\hompc}{\,h\,{\rm Mpc}^{-1}}

\newcommand{\mpcoh}{\,h^{-1}\,{\rm Mpc}}
\newcommand{\mpcohthree}{\,h^{-3}\,{\rm Mpc^3}}

\def\be{\begin{equation}}
\def\ee{\end{equation}}
\def\ba{\begin{eqnarray}}
\def\ea{\end{eqnarray}}

\def\nn{\nonumber}

\shorttitle{Cross-correlation between galaxy and 21cm}
\shortauthors{Song et al.}

\begin{document}
\title{Cosmological implication of cross--correlation between galaxy clustering and 21-cm line intensity mapping}
\correspondingauthor{Kyungjin Ahn}
\author{Yong-Seon Song}
\affiliation{Korea Astronomy and Space Science Institute, Daejeon 34055, Republic of Korea; ysong@kasi.re.kr}
\author{Minji Oh}
\affiliation{Department of Earth Sciences, Chosun University, Gwangju 61452, Republic of Korea}
\author{Kyungjin Ahn}
\affiliation{Department of Earth Sciences, Chosun University, Gwangju 61452, Republic of Korea; kjahn@chosun.ac.kr}
\author{Feng Shi}
\affiliation{School of Aerospace Science and Technology, Xidian University, Xi’an,
710126, China}

\begin{abstract}
The apparent anisotropies of galaxy clustering and 21-cm mapping in redshift space offer a unique opportunity to simultaneously probe cosmic expansion and gravity on cosmological scales through the Alcock--Paczynski (AP) effect and redshift-space distortions (RSD). Although improved theoretical models exist for anisotropic clustering, their applicability is limited by the non-perturbative smearing effect caused by the randomness of relative velocities. Here, we consider an alternative approach using the statistical power of cross-correlation between galaxy clustering and 21-cm line intensity mapping. Based on Fisher matrix analysis, fully incorporating nonlinear RSD, we estimate the benefit of combining both observables. We find that, for spectroscopy surveys like DESI combined with 21-cm line-intensity mapping surveys, constraints on the growth of structure and the cosmic expansion rate are improved by a factor of two relative to the galaxy auto-correlation. Crucially, such an observation can strongly constrain the neutral hydrogen (HI) content $\Omega_{\HI}$ to a sub-percent level. This level of precision unlocks the potential of this method to probe post-reionization astrophysics with enhanced precision. It would far surpass existing constraints from stacked 21-cm emission and break the degeneracy between $\Omega_{\HI}$ and the HI bias $b_{\HI}$ inherent in the linear-regime power-spectrum analysis. This cross-correlation approach effectively compensates for the loss of constraining power when using galaxy clustering alone.
\end{abstract}
 


\section{introduction}

In our current understanding of the universe, an unknown substance called dark matter dominates over the standard model particles at the present epoch. Despite many theoretical and observational efforts, the origin of dark matter is not yet clarified. Also, the existence of dark energy, which is supposed to drive the cosmic acceleration, indicates our incomplete understanding of the gravity on cosmological scales. It may imply modifications to Einstein's  theory of General Relativity.  A further insight into the origin and nature of dark energy or validity of general relativity is essential, and this is one of the primary goals in next-generation cosmology. 

The large-scale structure offers an opportunity to probe these issues by looking at the anisotropic galaxy clustering in redshift space. The observed galaxy distribution via the spectroscopic measurements is apparently distorted due to the peculiar velocity of galaxies along the line-of-sight direction, referred to as the redshift-space distortions (RSD). While RSD complicates the interpretation of the small-scale galaxy clustering, on large scales, the strength of anisotropies is simply characterized by the linear growth rate $f=d\ln G_\delta/d\ln a$, providing us a unique opportunity to probe the growth of structure, where $G_\delta$ and $a$ are the density growth function and scale factor of the Universe, respectively. On the other hand, the large-scale galaxy clustering data imprints a fossil record of the primeval baryon-photon fluid around the last-scattering surface, known as the baryon acoustic oscillations (BAO). The characteristic scale of  BAO can be used as a standard ruler, which enables us to determine the geometric distances of high-$z$ galaxies with greater precision. The key point to determine the geometric distances is to measure the clustering anisotropies over the BAO scales. Notice the fact that the anisotropies of the clustering pattern also arises from the apparent mismatch of the underlying cosmological model when we convert the redshift and angular position of each galaxy to the co-moving radial and transverse distances. This is the so-called Alcock-Paczynski (AP) effect, and with a prior knowledge of the characteristic scale of the BAO, the Hubble parameter $H(z)$ and angular diameter distance $D_A(z)$ of the high-$z$ galaxies can be separately measured. Thus, the anisotropic galaxy clustering can serve as a dual cosmological probe from which we can explore the origin of cosmic acceleration from the viewpoint of both dark energy and modification of gravity. A recent claim that the flat wCDM model prefers the dark energy equation of state $w\ne -1$ comes from both the measure of BAO \citep{DESIBAO} and the AP test \citep{Dong2023}, showing the importance of these effects in cosmology.

To simultaneously constrain cosmic expansion and structure growth, a detailed theoretical model of anisotropic clustering is essential. Although our focus is near the linear regime, small, non-negligible nonlinear systematics, including gravitational clustering, must be corrected. Perturbation theory (hereafter PT) is a popular template for this correction beyond linear theory, but its applicability is limited to the weakly nonlinear regime. Furthermore, RSD introduces non-trivial crosstalk between scales. Specifically, the Fingers-of-God (hereafter FoG) effect, caused by the virialized random motion of galaxies within halos, significantly suppresses the clustering amplitude along the line of sight, even on large scales. Since perturbation theory cannot fully describe the FoG effect, a phenomenological description must be introduced. Consequently, to avoid systematics, cosmological data analysis is restricted to large scales ($k \lesssim 0.1\,h\,\text{Mpc}^{-1}$), which severely reduces the statistical power to constrain cosmology.

In the pursuit of extracting maximal cosmological information exploiting the benefit of cross--correlation between galaxy clustering and 21-cm line intensity mapping (LIM), we present the combined results of the galaxy survey and 21-cm experiment to constrain the geometric distances and growth of structure. To be specific, we consider an accessible spectroscopic survey like Dark Energy Spectroscopic Instrument (hereafter DESI), which would be the best suited to probe the cosmic acceleration around $z\sim1$. Considering this survey setup, we discuss the impact of the FoG effect on the estimation of cosmological parameters, which has not been explored to the nonlinear order as far as we know. We show that while the uncertainty of the FoG effect in power spectrum is mostly degenerate with the coherent motion as a probe of the growth of structure, the combination of both power spectra breaks this degeneracy, thus improving the measurement accuracy of the coherent motion by a factor of two at most. As for the constraints on geometric distances, substantial improvement is not found.  

Another benefit of our nonlinear formalism for RSD in the cross--power spectrum is a potential to estimate the global, neutral hydrogen (hereafter HI) mass content in units of the critical density $\rho_{\rm crit}$, $\Omega_{\HI}\equiv \rho_{\HI}/\rho_{\rm crit}$, to a very high accuracy as we will show.  21-cm observations so far do not yet seem to have reached sensitivity high enough to measure $\Omega_{\HI}$ to a high accuracy. Constraints on $\Omega_{\HI}$ at $z\lesssim 1$ come mainly from two categories of observations: (1) stacked signal of 21-cm emission lines from individual galaxies \citep{Delhaize2013, Rhee2013, Rhee2018, Tramonte2020, DINGO2021} and (2) the clustering properties (e.g. power spectrum) of 21-cm line intensity maps \citep{Pen2009, Chang2010, Masui2013, Anderson2018, Li2021, Wolz2022, MeerKLASS2023, CHIME2023, CHIME2025}. For example, DINGO-VLA (Deep Investigations of Neutral Gas Origins--Very Large Array) team stacked 21-cm emission signals from thousands of galaxies, and have obtained $\Omega_{\HI}(z=0.05)=(0.38 \pm 0.04) \times 10^{-3}$ \citep{DINGO2021}. Even though the 1-$\sigma$ uncertainty is small, the theoretical uncertainty and the cosmic variance surpass the given uncertainty because the selection bias (only bright galaxies) may miss the contribution from faint galaxies and the sky coverage is still too small ($\sim 20\, {\rm deg}^2$) to guarantee the mitigation of the cosmic variance. Canadian Hydrogen Intensity Mapping Experiment (CHIME) stacked 21-cm emission signals from individual objects but intentionally including nearly objects as well, to yield an estimate of the HI clustering property:  \( A_{\HI}(z=0.84) \equiv 10^3 \, \Omega_{\HI} \left( b_{\HI} + \langle f \mu^2 \rangle \right)
= 1.51^{+3.60}_{-0.97} \) for Luminous Red Galaxy (LRG), \( A_{\HI}(z=0.96) = 6.76^{+9.04}_{-3.79} \) for Emission Line Galaxy (ELG), and \( A_{\HI}(z=1.20) = 1.68^{+1.10}_{-0.67} \) for QSOs, showing the dependence of $A_{\HI}$ on the type of galaxies \citep{CHIME2023}, where $b_{\HI}$ is the HI linear bias, $f$ is the growth rate of structure, and $\mu$ represents the cosine of the angle between the comoving wavevector $\bfk$ and the line-of-sight.

We will show that the usual degeneracy in estimation of $\Omega_{\HI}$ and $b_{\HI}$, shown in the clustering analysis by e.g. CHIME (\citealt{CHIME2023}; see also the one by MeerKAT+WiggleZ: \citealt{MeerKLASS2023}), can be broken by our formalism.
While the full compilation of CHIME data may enable the estimation of the cross power spectrum with galaxy survey data, the standard linear-regime assumption in power spectrum analysis is insufficient to break the degeneracy between $\Omega_{\HI}$ and $b_{\HI}$. Furthermore, although a recent study \citep{CHIME2025} successfully reported the first highly significant detection of the HI auto power spectrum (yielding $A_{\HI}(z = 1.16) = 2.59^{+1.26}_{-0.78}\text{(stat.)}^{+2.45}_{-0.47}\text{(sys.)}$, where ``stat.'' and ``sys.'' denote statistical and systematic uncertainties, respectively), their methodology excludes data at $k < 0.4\,h/\text{Mpc}$ presumably due to foreground avoidance (a scheme to exclude $k$ values that is most prone to the spectrally smooth foreground). The largeness of $k$ in such cases requires a more sophisticated theoretical treatment to model the non-linear regime. Alternatively, efficient foreground-removal methods might recover large-scale modes otherwise inaccessible. Recent simulation-based studies have demonstrated substantial progress in recovering the HI auto- and cross-power spectra on large scales, including instrumental beam effects \citep[e.g.,][]{Shi2024,Wang2026}.
Our formalism presented in this paper, which focuses on the quasi-linear regime $k\la 0.2 \,h^{-1} \rm Mpc$ at $z\sim 1$ and if realized through observations, is expected to provide a much more accurate estimate of $\Omega_{\HI}$ than these studies and thus improve its measurement significantly (Sec.~\ref{sec:xHI}), with the degeneracy between $\Omega_{\HI}$ and $b_{\HI}$ broken.

Our analysis adopts a highly generalized approach regarding both the underlying cosmological model and the impact of radio foregrounds. By remaining fundamentally model-independent, we avoid assuming a specific framework, such as the $\Lambda$CDM universe, and explicitly account for residual signals caused by imperfect radio foreground removal. Although multi-tracer cosmology using galaxies and HI emission has been studied extensively, the typical approach is to fix the cosmological model, neglect the radio foreground, or both. For example, \citet{Berti2024} investigated the potential of joint galaxy and HI analyses, finding that HI observations by the SKA can provide constraining power on cosmological parameters comparable to existing galaxy surveys. However, their study assumed a $\Lambda$CDM cosmology and neglected radio foregrounds. Consequently, our work serves as a more robust and realistic forecast for future joint HI and galaxy analyses.

This paper is organized as follows. In Section~\ref{sec:method}, we lay out a general formalism for auto- and cross-correlation power spectrums of the galaxy survey and the 21-cm LIM, but with a careful treatment of RSD extended to the non-linear regime. In Section~\ref{sec:result}, we show the expected constraints of cosmological and astrophysical parameters from our Fisher matrix analysis, focusing on the improved constraint on dynamics-related cosmological parameters and $\Omega_{\rm HI}$. In Section~\ref{sec:conclusion}, we discuss the prospect of our methodology.

\section{methodology}
\label{sec:method}

\subsection{The power spectrum in redshift space}
\label{subsec:hybrid_DM_RSD_cross}

The anisotropic power spectrum in redshift space is given as
\begin{equation}
P_{ab}(k,\mu)= \mathscr{C}_{a}\mathscr{C}_{b}\int d^3\bfx\,e^{i\,\bfk\cdot\bfx}
\bigl\langle e^{j_1A_1}A_2^aA_3^b\bigr\rangle\,,
\label{eq:Pkred_exact}
\end{equation}
where $a$ and $b$ denote either the galaxy ($g$) or the hydrogen line ($\rm H$), with normalization factors $\mathscr{C}_{g}=1$ and $\mathscr{C}_{\rm H}=p_{S}$ (we defer the definition of $A_1$, $A_2$ and $A_3$ to Eq.~\ref{eq:As}). Eq.~(\ref{eq:Pkred_exact}) can be used both for auto-correlation and cross-correlation power spectra. As for the hydrogen line, $p_S$ is the sky-averaged power received by a radio antenna in the frequency bandwidth $\Delta \nu$ \citep{2010ApJ...721..164S}:
\ba
p_S=p_S(z)=k_B\hat{T}_{\rm sig} \Delta \nu
\label{eq:ps_antenna1}
\ea
where $\hat{T}_{\rm sig}$, the sky-averaged and observed temperature of the 21-cm background originating from redshift $z$\footnote{The observing frequency $\nu$ is related to $z$ implicitly by $\nu=1.42/(1+z)\,{\rm GHz}$},  is given by
\ba
\hat{T}_{\rm sig}(z) = 188\frac{x_{\rm HI}(z)\Omega_{\rm H,0} (1+z)^2 h}{H(z)/H_0} {\rm mK}.
\label{eq:Tsig}
\ea
In Eq.~\ref{eq:Tsig}, $\Omega_{\rm H,0}$ represents the present-day hydrogen mass fraction defined as $\Omega_{\rm H,0} = X\Omega_{b,0} = 0.03753$. Here, $\Omega_{b,0} \equiv \rho_{b,0}/\rho_{\rm crit,0}$ is the cosmic baryon density in units of the critical density at present and $X \equiv \rho_{\rm H}/\rho_b$ is the hydrogen mass fraction in baryon. The neutral hydrogen fraction, defined as $x_{\rm HI}(z) \equiv \rho_{\rm HI}(z)/\rho_{\rm H}(z)$, is treated as an unknown parameter with a fiducial value of $x_{\rm HI} = 0.027$ at redshift $z \simeq 1$. This fiducial value is to reflect those estimated through Lyman-alpha forest observations, ranging from $\Omega_{\rm HI}(z)\equiv x_{\rm HI}(z) \Omega_{\rm H,0} = (0.7\pm 0.3)\times 10^{-3}$ \citep{Rao2006} to $\Omega_{\rm HI}(z)= (1.2\pm 0.3)\times 10^{-3}$ \citep{Grasha2020} at $z\simeq 1$. 

The mapping between real space $\bfr$ and redshift space $\bfs$ is given by
\beq
\label{eq:mapping}
\bfs=\bfr+\frac{\bfv \cdot \hat{z}}{aH}\hat{z},
\eeq
where $\bfv$, $a$, $H$, and $\hat{z}$ represent the peculiar velocity, the scale factor of the expansion, the Hubble parameter, and the unit vector toward the line of sight, respectively. The redshift distortion effect (RSD hereafter) is exhibited in this power spectrum formulation, causing an apparent anisotropy to the measured power spectrum. The observer is assumed to be located far away from the targeted galaxies. The variable $j_1$ and fields $A_i$ in Eq.~(\ref{eq:Pkred_exact}) are defined, with $\mu=\cos{\mathbf{k}\cdot\hat{z}}$, as
\begin{eqnarray}
&j_1= -i\,k\mu ,\nonumber\\
&A_1=u_z(\bfr)-u_z(\bfr'),\nonumber\\
&A_2^a=\Bigl[\delta_a(\bfr)+\,\nabla_zu_z(\bfr)\Bigr],\nonumber\\
&A_3^a=\Bigl[\delta_a(\bfr')+\,\nabla_zu_z(\bfr')\Bigr].
\label{eq:As}
\end{eqnarray}

The quantities $\bfx$ and $\bfu$ are given by $\bfx=\bfr-\bfr'$ and $\bfu\equiv-\bfv/(aH)$. The function $u_z$ is the line-of-sight  component of $\bfu$ and the density contrast of a tracer $a$, $\delta_a$, is written in terms of the tracer-specific biases $b_{a}$ and the matter overdensity $\delta$. Here we adopt the equivalence principle that the velocities are consistent among different tracers. The velocity fields are assumed to be those after correcting the possible density bias, as the surveys of presumed velocity fields are in fact probing the momentum fields \citep{Appleby2023} and thus affected by the density bias.

The full expression for the RSD power spectrum is split into perturbative and non--perturbative terms for practical purpose. In the quasi non--linear regime, Gaussian statistics are applied to express both density and velocity fluctuations in infinite series of perturbative expansions. The non--perturbative part accounting for FoG is assumed to be separable from the perturbative part, and will be represented by an overall multiplication factor $D^{\rm FoG}$. The full anisotropic power spectrum (Eq.~\ref{eq:Pkred_exact}) is expressed approximately as
\begin{eqnarray}
&&P_{ab}(k,\mu)
=\mathscr{C}_{a}\mathscr{C}_{b} D^{\rm FoG}(k\mu \sigma_p)\Bigl[P_{\delta_a\delta_b}(k) \nn \\
&&+\mu^2P_{\delta_a\Theta}(k) +\mu^2P_{\delta_b\Theta}(k)+\mu^4P_{\Theta\Theta}(k)+A_{ab}(k,\mu) \nn \\
&&+B_{ab}(k,\mu)+T_{ab}(k,\mu)+F_{ab}(k,\mu)\Bigr],
\label{eq:Pkred_final}
\end{eqnarray}
where $\Theta\equiv -\nabla\cdot \bfv/(aH)=\nabla \cdot \bfu$, and $\sigma_p$ is the line-of-sight velocity variance $\langle u_z^2 \rangle_c$ that we take as a free parameter in this paper. Here, $P_{\delta_a\delta_b}$, $P_{\Theta\Theta}$ and $P_{\delta_a\Theta}$ are power spectra of cross-correlation between the overdensities $\delta_a$ and $\delta_b$, of the velocity-divergence $\Theta$, and of the cross-correlation between $\delta_a$ and $\Theta$, respectively (see below why $\Theta$ is independent of the tracer species). The simple expansion in $\mu^2$ and $\mu^4$ in Eq.~(\ref{eq:Pkred_final}) is possible when using the velocity-modulated density fields $A_2$ and $A_3$ in Eq.~(\ref{eq:Pkred_exact}) to account for RSD, under the assumption of the irrotational flow.

We assume that velocity fields of any type of tracers are identical to each other, or there is no specific velocity bias, and therefore the velocity divergence $\Theta$ (subscripted) in Eq.~(\ref{eq:Pkred_final}) is blind to the tracer type (subscripts $a$ and $b$).
Note also that the factorized term $D^{\rm FoG}$ represents a non-perturbative contribution coming from the zero-lag correlation of velocity fields and plays a role in suppressing the overall amplitude at small scales. We adopt a Gaussian functional form for $D^{\rm FoG}$: $D^{\rm FoG}(k\mu\sigma_{p})=\exp{\large[-(k\mu\sigma_p)^2]}$. For more details of the terms and formalism, see Section~2.1 of \citet{Song_2018}\footnote{We have extended the original formalism by \citet{Song_2018} into a double-tracer case, or the galaxy and the HI 21-cm, in Section~\ref{sec:gHcross} by employing the fact that $A_1$ (Eq.~\ref{eq:As}) is a common quantity among different tracers. This reflects the fact that a galaxy, whether traced by its optical emission or the HI emission, has a single-valued peculiar velocity.}. The observed power spectrum $\tilde{P}_{ab}(k,\mu)$ will suffer from observational uncertainties, which will be specified in terms of a covariance matrix in Section~\ref{subsec:fisher}.

High-order power spectra  in Eq.~(\ref{eq:Pkred_final}), necessary to account for the nonlinearity at relatively large $k$ ($k\gtrsim 0.1\,h\,\text{Mpc}^{-1}$), are constructed from products of $A_1$, $A_2$ and $A_3$ from Eq.~(\ref{eq:As}):
\begin{eqnarray}
  A_{ab}(k,\mu)&=& j_1\,\int d^3\bfx \,\,e^{i\bfk\cdot\bfx}\,\,\langle A_1A_2^aA_3^b\rangle_c,\nonumber\\
  B_{ab}(k,\mu)&=& j_1^2\,\int d^3\bfx \,\,e^{i\bfk\cdot\bfx}\,\,\langle A_1A_2^a\rangle_c\,\langle A_1A_3^b\rangle_c,\nonumber\\
  T_{ab}(k,\mu)&=& \frac{1}{2} j_1^2\,\int d^3\bfx \,\,e^{i\bfk\cdot\bfx}\,\,\langle A_1^2A_2^aA_3^b\rangle_c,\nonumber \\
  F_{ab}(k,\mu)&=& -j_1^2\,\int d^3\bfx \,\,e^{i\bfk\cdot\bfx}\,\,\langle u_z u_z'\rangle_c\langle A_2^aA_3^b\rangle_c,
\label{eq:ABTF}
\end{eqnarray}
where $u_z(\bfr)$ and $u_z(\bfr')$ are abbreviated as $u_z$ and $u'_z$, respectively, and this convention is used hereafter. 
$\langle F_1 F_2 \dots F_n\rangle_c$ is the ``connected'' average of quantities $\{F_1,\,F_2,\,\dots,F_n\}$, such that e.g.
\begin{eqnarray}
\langle F_1 F_2 \rangle_c &=& \langle F_1 F_2 \rangle - \langle F_1 \rangle\langle F_2 \rangle , \nn \\
\langle F_1 F_2 F_3 \rangle_c &=& \langle F_1 F_2 F_3 \rangle - \langle F_1 F_2 \rangle\langle F_3 \rangle - \langle F_1 F_3 \rangle\langle F_2 \rangle \nn \\
&&- \langle F_2 F_3 \rangle\langle F_1 \rangle + 2\langle F_1 \rangle\langle F_2 \rangle\langle F_3 \rangle .
\label{eq:connected_average}
\end{eqnarray}
Even though not shown explicitly, all these high-order power spectra ($A_{ab}$, $B_{ab}$, $T_{ab}$, $F_{ab}$) in Eq.~(\ref{eq:ABTF}) are real-valued quantities. For the detailed discussion and derivation, refer to \citet{Song_2018} and \citet{Taruya:2013my}.



The observed power spectra for given tracers (galaxy and HI emission in our case) in the weakly non--linear regime are expressed as a combination of tracer-specific bias parameters and the underlying matter fluctuation quantities, namely the matter overdensity $\delta$ and the matter velocity divergence $\Theta$. This formalism includes both local and non--local contributions. In this paper, we adopt the prescription proposed by \citet{McDonald_bias1} and \citet{Saito2014}, which has been applied to the Baryon Oscillation Spectroscopic Survey (BOSS) in Sloan Digital Sky Survey (SDSS) \citep{2011AJ....142...72E, 2013AJ....145...10D}. We adaptively use this prescription for our two main tracers, galaxy and HI emission. Apart from stochastic terms, this formalism is the ``hybrid RSD model'' we developed elsewhere to estimate cosmological parameters accurately from the auto-power spectrum of the galaxy clustering \citep{10.1093/mnras/stae2383}. This hybrid RSD model is composed of a perturbation theory valid up to the two-loop order and a scaling law for even higher-order loop corrections to the power spectrum \citep{Song_2018}. We adopt a limiting wavenumber $k_{\rm max}\simeq {\bf 0.2}h\,{\rm Mpc}^{-1}$, where the nontrivial contribution from high-order loop corrections are substantial enough to break the degeneracy between the linear bias and the pivot-scale variance such as $\sigma_8$. Our extension of this formalism to HI surveys will be found to also break the degeneracy between $\Omega_{\rm HI}$ and $b_{\rm HI}$.

\subsection{Galaxy auto--correlation}
\label{sec:ggauto}

We first link power spectra traced by galaxies to the underlying matter-perturbation power spectra in the leading order. The leading-order contribution to the galaxy auto-correlation power spectrum $P_{gg}$ (Eq.~\ref{eq:Pkred_final} with $a=b=g$) comes from $P_{\delta_g\delta_g}$, $P_{\delta_g \Theta}$ and $\Ptt$, and the objective is to express these in terms of the matter power spectra $\Pdd$, $\Pdt$ and $\Ptt$. First, the galaxy number-density power spectrum $P_{\delta_g\delta_g}$, is given by
\ba
P_{\delta_g\delta_g}(k)&&=b_{g1}^2P_{\delta\delta}(k)+2b_{g1}b_{g2}P_{b2,\delta}(k)+2b_{g1}b_{gs2}P_{bs2,\delta}(k) \nn \\
&&+2b_{g1}b_{g3\rm{nl}}\sigma_3^2(k)P^{\rm{L}}_{\rm m}(k)+b_{g2}^2P_{b22}(k)\nn \\
&& +2b_{g2}b_{gs2}P_{b2s2}(k) +b_{gs2}^2P_{bs22}(k), 
\label{eq:pgg_bias}
\ea
where $P_{\delta\delta}$ and $P_{\rm m}^{\rm L}$ are the non-linear and linear matter power spectra in real space, respectively, and the non-local bias parameters $b_{gs2}$ and $b_{g3\rm{nl}}$ are approximately related to the linear bias $b_{g1}$ at first order and therefore our observational forecast will \emph{exclude} these non-local bias parameters (see Section~\ref{subsec:fisher}). They are given as $b_{gs2}=-\frac{4}{7}(b_{g1}-1)$ and $b_{g3\rm{nl}}=\frac{32}{315}(b_{g1}-1)$ \citep{Baldauf:2012hs, 1994ApJ...433....1B, 2014PhRvD..90l3522S,PhysRevD.85.083509}. The rest of the terms in Eq.~(\ref{eq:pgg_bias}), $P_{b2,\delta}$, $P_{bs2,\delta}$, $\sigma_3^2$, $P_{b22}$, $P_{b2s2}$, and $P_{bs22}$ are one-loop corrections, but expressible in $P_{\rm m}^{\rm L}$, as can be seen in \citet{Hector_bias} and \citet{Saito2014}.
Second, the cross-correlation power spectrum of galaxy-density $\delta_g$ and velocity-divergence $\Theta$, $P_{\delta_g\Theta}$, is given by
\ba
P_{\delta_g\Theta}(k)&&=b_{g1}P_{\delta\Theta}(k)+b_{g2}P_{b2,\Theta}(k)+b_{gs2}P_{bs2,\Theta}(k)\nn
\\
&&+b_{g3\rm{nl}}\sigma_3^2(k)P^{\rm{L}}_{\rm m}(k)\,,
\label{eq:pgt_bias}
\ea
where $P_{\delta_g\Theta}$ is the $\delta_g$--$\Theta$ cross-correlation power spectrum in real space, and the higher-order power spectra $P_{b2,\Theta}$ and $P_{bs2,\Theta}$ are again given by one--loop integrals as shown in \citet{Hector_bias} and \citet{Saito2014}. Third, as we assume unbiaed velocity tracers, the auto-correlation power spectrum of the galaxy velocity-divergence $\Theta_g$ is simply identical to $\Ptt$.

So far, the leading order galactic power spectra $P_{\delta_g \delta_g}$, $P_{\delta_g \Theta_g}$ and $P_{\Theta_g \Theta_g}$ that compose $P_{gg}$ have been expressed in terms of the matter power spectra $\Pdd$, $\Pdt$ and $\Ptt$ in the leading order and other higher-order power spectra. These leading-order matter power spectra are nonlinear, and thus are not given in terms of the linear power spectrum ad hoc. We therefore adopt the ``hybrid RSD model'', a perturbative scheme combining analytical formalism with N-body simulation results, to calculate $\Pdd$, $\Pdt$ and $\Ptt$ (See Appendix \ref{appendix:hybrid_DM_RSD} for details). This hybrid scheme has the following procedures: (1) take a ``fiducial'' cosmology with a specific set of cosmological parameters (any quantity $A$ in the fiducial cosmology to be denoted by $\bar{A}$), (2) use a perturbation theory to calculate the power spectrum accurate up to two--loop order (to be denoted by the superscript ``th'' to represent the ``theory''), (3) perform an N-body simulation, (4) compare the results of steps (2) and (3) to find the difference (to be denoted by the superscript ``res'' to represent the ``residual''), which occurs beyond some wavenumber $k$ where the perturbation theory breaks down, and finally (5) apply a transformation rule for $P^{\rm th}$ and $P^{\rm res}$ using the density growth function $G_\delta$ and the velocity growth function $G_\Theta$ for a cosmology different from the fiducial one. We use fiducial growth functions $\bar{G}_\delta$ and $\bar{G}_\Theta$ calculated by our own Boltzmann solver. Step (5) is crucial when we perform the Fisher analysis, because cosmological parameters need to be varied for the analysis. Without such a transformation rule, running a new N-body simulation every time a parameter is altered becomes too expensive computationally.

$A$, $B$, $T$ and $F$ in Eq.~(\ref{eq:Pkred_final}) are higher-order corrections showing explicit anisotropy. It is sufficient to keep only the leading-order term in bias for these corrections, and thus we only apply the linear bias $b_{g1}$ when using Eqs.~(\ref{eq:As}) and (\ref{eq:ABTF}), and again assume an unbiased velocity field. First, $A_{gg}(k,\mu)$ under a certain cosmology can be estimated from a fiducial cosmology by the following transformation rules (similar to step (5) in the hybrid RSD model described above)
\ba
\label{eq:estimatedAn}
&&A_{gg}(k,\mu)
= \sum_{n=1}^{4} {\cal A}_{n,gg}   \nn \\
&&=\left(G_\delta/\bar G_\delta\right)^2\left(G_\Theta/\bar G_\Theta\right)
\bar{\cal A}_{1,gg}
+\left(G_\delta/\bar G_\delta\right)\left(G_\Theta/\bar G_\Theta\right)^2
\bar{\cal A}_{2,gg}\nn\\
&&+\left(G_\delta/\bar G_\delta\right)\left(G_\Theta/\bar G_\Theta\right)^2
\bar{\cal A}_{3,gg}
+\left(G_\Theta/\bar G_\Theta\right)^3
\bar{\cal A}_{4,gg},
\ea
where the barred quantities $\bar{\cal A}_{n,gg}$, defined in Appendix \ref{appendix:hybrid_cross}, are measured in fiducial cosmology and they are rescaled to other cosmology using the density growth function $G_\delta$ and the velocity growth function $G_\Theta$. Note that $\bar{\cal A}_{n,gg}$'s have an explicit dependence on $b_{g1}$ except for $\bar{\cal A}_{4,gg}$, as shown in Eq.~(\ref{eq:Anab}).

Similarly, $B_{gg}(k,\mu)$, $T_{gg}(k,\mu)$, and $F_{gg}(k,\mu)$ can be expressed as
\bea
\label{eq:estimatedBn}
&&B_{gg}(k,\mu) = \sum_{n=1}^{3} {\cal B}_{n,gg}   \nn
\\
&&= \left(G_\delta/\bar G_\delta\right)^2\left(G_\Theta/\bar G_\Theta\right)^2
\bar{\cal B}_{1,gg}+\left(G_\delta/\bar G_\delta\right)\left(G_\Theta/\bar G_\Theta\right)^3
\bar{\cal B}_{2,gg} \nn \\
&&+\left(G_\Theta/\bar G_\Theta\right)^4
\bar{\cal B}_{3,gg},
\eea
\ba
\label{eq:estimatedTn}
&&T(k,\mu)= \sum_{n=1}^{7} {\cal T}_{n,gg}  \nn
\\
&&= \left(G_\delta/\bar G_\delta\right)^2\left(G_\Theta/\bar G_\Theta\right)^2
\bar{\cal T}_{1,gg}
+\left(G_\delta/\bar G_\delta\right)\left(G_\Theta/\bar G_\Theta\right)^3
\bar{\cal T}_{2,gg}\nn\\
&&+\left(G_\delta/\bar G_\delta\right)\left(G_\Theta/\bar G_\Theta\right)^3
\bar{\cal T}_{3,gg}
+\left(G_\Theta/\bar G_\Theta\right)^4
\bar{\cal T}_{4,gg}
\nn
\\
&&
+\left(G_\delta/\bar G_\delta\right)^2\left(G_\Theta/\bar G_\Theta\right)^2
\bar{\cal T}_{5,gg}
+\left(G_\delta/\bar G_\delta\right)\left(G_\Theta/\bar G_\Theta\right)^3
\bar{\cal T}_{6,gg}
\nn
\\
&&
+\left(G_\Theta/\bar G_\Theta\right)^4
\bar{\cal T}_{7,gg}, 
\ea
and,
\ba
\label{eq:estimatedFn}
&&F(k,\mu) = \sum_{n=1}^{3} {\cal F}_{n,gg}   \nn \\
&&= \left(G_\delta/\bar G_\delta\right)^2\left(G_\Theta/\bar G_\Theta\right)^2
\bar{\cal F}_{1,gg}+\left(G_\delta/\bar G_\delta\right)\left(G_\Theta/\bar G_\Theta\right)^3\bar{\cal F}_{2,gg} \nn \\
&&+\left(G_\Theta/\bar G_\Theta\right)^4
\bar{\cal F}_{3,gg}. 
\ea
The derivations of Eqs.~(\ref{eq:estimatedAn}-\ref{eq:estimatedFn}) are described by Eqs.~(\ref{eq:estimatedAn_ab}), (\ref{eq:estimatedBn_ab}), (\ref{eq:estimatedTn_ab}) and (\ref{eq:estimatedFn_ab}), respectively. The components $\bar{\cal B}_{n,gg}$, $\bar{\cal T}_{n,gg}$ and $\bar{\cal F}_{n,gg}$ are derived in Eqs.~(\ref{eq:Bnab}), (\ref{eq:Tnab}) and (\ref{eq:Fnab}), respectively. The accuracy of this treatment and its impact on the RSD model are discussed in \citet{Zheng19}, and allows a rather accurate estimation of $b_{g1}$ by breaking the degeneracy between $b_{g1}$ and $G_\delta$. This is due to the fact that our nonlinear formalism roughly encapsulates the power of the joint analysis of the linear power spectrum and the bi-spectrum suggested by \citet{2015JCAP...08..007S}.

The selected fiducial values of galaxy biases and the overall DESI sample descriptions are presented in Table~\ref{tab:DESI}. We use a limited galaxy samples: ELGs in the redshift range $z=1.0$--$1.2$.

\begin{deluxetable*}{cccccccc}
\tablewidth{0pt}
\tablecaption{DESI survey specification}
\tablehead{  \colhead{$z$} & \colhead{$n_{g}^{\rm LRG}$} & \colhead{$n_{g}^{\rm ELG}$} & \colhead{$V$} & \colhead{$b_1^{\rm LRG}$}  & \colhead{$b_1^{\rm ELG}$} & \colhead{$b_2^{\rm LRG}$}  & \colhead{$b_2^{\rm ELG}$}
}
  \startdata
  0.4--0.6 & $4.9\times 10^{-4}$ & $1.6\times 10^{-4}$ & 3.5 & 2.22 & 1.10 & 0.10 & 0.10\\
  0.6--0.8 & $9.9\times 10^{-4}$ & $1.4\times 10^{-3}$ & 5.4 & 2.45 & 1.21 & 0.67 & 0.67\\
  0.8--1.0 & $3.9\times 10^{-4}$ & $1.7\times 10^{-3}$ & 7.0 & 2.69 & 1.33 & 1.40 & 1.40\\
  1.0--1.2 & $2.4\times 10^{-5}$ & $9.8\times 10^{-4}$ & 8.4 & 2.94 & 1.45 & 2.48 & 2.48\\
  1.2--1.4 & --- & $5.8\times 10^{-4}$ & 9.4 &  --- & 1.57 & --- & 3.91\\
  1.4--1.6 & --- & $2.3\times 10^{-4}$ & 10. &  --- & 1.70 & --- & 5.68\\
\enddata

\label{tab:DESI}
\tablecomments{The expected number density of LRGs and ELGs are $n_g^{\rm LRG}$ ($h^3{\rm Mpc}^{-3}$) and $n_g^{\rm ELG}$ ($h^3{\rm Mpc}^{-3}$), respectively. $V$ ($h^{-3}\,{\rm Gpc}^3$) is the survey volume (comoving) at each redshift bin. For this study, we use ELG galaxies from a single redshift bin $z=1.0$--$1.2$. These specific values are taken from the blueprints of the DESI experiment \citep{DESI16I}. Fiducial values of bias for LRG and ELG are presented as well, calculated by the formulas $b_{1}^{\rm LRG}(z)D(z) = 1.7$ and $b_{1}^{\rm ELG}(z)D(z) = 0.84$ \citep{DESI16I}, with $D$ being the linear growth factor normalized to unity at present time, i.e., $D(z)=G_\delta(z)/G_\delta(0)$. For $b_2^{\rm LRG}$ and $b_2^{\rm ELG}$, we use the ``halofit'' results \citep{halofit2012}. The $\Lambda$CDM model with Planck-estimated \citep{PLANK2015} parameters is adopted as the fiducial cosmology. }
\end{deluxetable*}

\subsection{21-cm auto--correlation}
\label{sec:hhauto}
We follow the procedure that is identical to that in Section~\ref{sec:ggauto}, to calculate the HI auto-correlation power spectrum $P_{\Htr\Htr}$ (Eq.~\ref{eq:Pkred_final}) in terms of the matter power spectra and corresponding high-order correction terms.
The first leading-order power spectrum in Eq.~(\ref{eq:Pkred_final}) with $a=b=\Htr$, $P_{{\delta_{\Htr}\delta_{\Htr}}}$, is given by
\ba
P_{{\delta_{\Htr}\delta_{\Htr}}}(k)&=&b_{\Htr 1}^2P_{\delta\delta}(k)+2b_{\Htr 1}b_{\Htr 2}P_{b2,\delta}(k) 
\no
\\
&+&2b_{\Htr 1}b_{\Htr s2}P_{bs2,\delta}(k) 
\no
\\
&+&2b_{\Htr 1}b_{\Htr 3\rm{nl}}\sigma_3^2(k)P^{\rm{L}}_{\rm m}(k)+b_{\Htr 2}^2P_{b22}(k) \no
\\
&+&2b_{\Htr 2}b_{\Htr s2}P_{b2s2}(k)
+b_{\Htr s2}^2P_{bs22}(k)
\ea
where the non-local bias terms are related to the linear bias term at first order by $b_{\Htr s2}=-\frac{4}{7}(b_{\Htr 1}-1)$ and $b_{\Htr 3\rm{nl}}=\frac{32}{315}(b_{\Htr 1}-1)$ \citep{Baldauf:2012hs, 1994ApJ...433....1B, 2014PhRvD..90l3522S,PhysRevD.85.083509}.

The power spectrum of the HI density $\delta_{\Htr}$ and the HI velocity divergence $\Theta_{\Htr} $ ($=\Theta$) is given by,
\ba
P_{\delta_{\Htr}\Theta}(k)&&=b_{\Htr 1}P_{\delta\Theta}(k)+b_{\Htr 2}P_{b2,\Theta}(k)+b_{\Htr s2}P_{bs2,\Theta}(k)\nn \\
&&+b_{\Htr 3\rm{nl}}\sigma_3^2(k)P^{\rm{L}}_{\rm m}(k)\,.
\label{eq:pHt_bias}
\ea
And, in the absence of velocity bias, the auto-power spectrum of $\Theta_{\Htr}$ is again identical to $P_{\Theta\Theta}$, just as in the case of $\Theta_g$ (Section~\ref{sec:ggauto}).
The higher--order terms $A_{\Htr\Htr}(k,\mu)$, $B_{\Htr\Htr}(k,\mu)$, $T_{\Htr\Htr}(k,\mu)$, and $F_{\Htr\Htr}(k,\mu)$ for $P_{\Htr\Htr}$ follow the transformation rules shown in Eqs.~(\ref{eq:estimatedAn}-\ref{eq:estimatedFn}), with a simple replacement of the bias paramter $b_{g1}$ by $b_{\Htr 1}$. We of course use the dimensional normalization coefficient $\mathscr{C}_{\rm H}=p_S$ as described in Section~\ref{subsec:hybrid_DM_RSD_cross}. The general expression can be found from Eqs.~(\ref{eq:estimatedAn_ab}-\ref{eq:estimatedFn_ab}) in Appendix \ref{appendix:hybrid_cross}.

\subsection{Galaxy--21-cm cross--correlation}
\label{sec:gHcross}
The cross--power spectrum $P_{g \rm H}$ between galaxy and HI, given in Eq.~(\ref{eq:Pkred_final}) with ($a$, $b$)=($g$, $\Htr$), have the following leading-order power spectra expressed in the matter power spectra as
\ba
P_{\delta_{g}\delta_{\Htr}}(k)&&=b_{g1}b_{\Htr 1}P_{\delta\delta}(k) \nn \\
&& +b_{g1}b_{\Htr 2}P_{b2,\delta}(k)+b_{\Htr 1}b_{g2}P_{b2,\delta}(k) \no \\
&& +b_{g1}b_{\Htr s2}P_{bs2,\delta}(k)+b_{\Htr 1}b_{gs2}P_{bs2,\delta}(k) \no \\
&& +b_{g1}b_{\Htr 3\rm{nl}}\sigma_3^2(k)P^{\rm{L}}_{\rm m}(k)+b_{\Htr 1}b_{g3\rm{nl}}\sigma_3^2(k)P^{\rm{L}}_{\rm m}(k) \no \\
&& +b_{g2}b_{\Htr s2}P_{b2s2}(k)+b_{\Htr 2}b_{gs2}P_{b2s2}(k)\no \\
&& +b_{g2}b_{\Htr 2}P_{b22}(k)+b_{gs2}b_{\Htr s2}P_{b22}(k), 
\ea
\ba
P_{\delta_{\Htr}\Theta}(k) &&= b_{\Htr 1}P_{\delta\Theta}(k)+b_{\Htr 2}P_{b2,\Theta}(k)+b_{\Htr s2}P_{bs2,\Theta}(k) \nn\\
&& + b_{\Htr 3\rm{nl}}\sigma_3^2(k)P^{\rm{L}}_{\rm m}(k),
\ea
\ba
P_{\delta_{g}\Theta}(k) &&= b_{g1}P_{\delta\Theta}(k)+b_{g2}P_{b2,\Theta}(k)+b_{gs2}P_{bs2,\Theta}(k) \nn \\
&& + b_{g3\rm{nl}}\sigma_3^2(k)P^{\rm{L}}_{\rm m}(k)\,,
\label{eq:gH_pgt_bias}
\ea
and again $P_{\Theta_g \Theta_{\rm H}}=P_{\Theta\Theta}$ in the absence of the velocity bias.

The higher order correction terms for $P_{g \rm H}$, namely $A_{g\Htr}(k,\mu)$, $B_{g\Htr}(k,\mu)$, $T_{g\Htr}(k,\mu)$, and $F_{g\Htr}(k,\mu)$, are again scaled from the fiducial cosmology (barred quantities) following the hybrid RSD model (Section~\ref{sec:ggauto}) as
\ba
\label{eq:estimatedAn_gH}
&&A_{g\Htr}(k,\mu)
= \sum_{n=1}^{4} {\cal A}_{n,g\Htr}   \nn \\
&&=\left(G_\delta/\bar G_\delta\right)^2\left(G_\Theta/\bar G_\Theta\right)
\bar{\cal A}_{1,g\Htr}
+\left(G_\delta/\bar G_\delta\right)\left(G_\Theta/\bar G_\Theta\right)^2
\bar{\cal A}_{2,g\Htr}\nn\\
&&+\left(G_\delta/\bar G_\delta\right)\left(G_\Theta/\bar G_\Theta\right)^2
\bar{\cal A}_{3,g\Htr}
+\left(G_\Theta/\bar G_\Theta\right)^3
\bar{\cal A}_{4,g\Htr},
\ea
\ba
\label{eq:estimatedBn_gH}
&&B_{g\Htr}(k,\mu) = \sum_{n=1}^{3} {\cal B}_{n,g\Htr}   \nn
\\
&&= \left(G_\delta/\bar G_\delta\right)^2\left(G_\Theta/\bar G_\Theta\right)^2
\bar{\cal B}_{1,g\Htr}+\left(G_\delta/\bar G_\delta\right)\left(G_\Theta/\bar G_\Theta\right)^3 \bar{\cal B}_{2,g\Htr} \nn \\
&&+\left(G_\Theta/\bar G_\Theta\right)^4
\bar{\cal B}_{3,g\Htr},
\ea

\ba
\label{eq:estimatedTn_gH}
&&T_{g\Htr}(k,\mu)= \sum_{n=1}^{7} {\cal T}_{n,g\Htr}  \nn
\\
&&= \left(G_\delta/\bar G_\delta\right)^2\left(G_\Theta/\bar G_\Theta\right)^2
\bar{\cal T}_{1,g\Htr}
+\left(G_\delta/\bar G_\delta\right)\left(G_\Theta/\bar G_\Theta\right)^3
\bar{\cal T}_{2,g\Htr}\nn\\
&&+\left(G_\delta/\bar G_\delta\right)\left(G_\Theta/\bar G_\Theta\right)^3
\bar{\cal T}_{3,g\Htr}
+\left(G_\Theta/\bar G_\Theta\right)^4
\bar{\cal T}_{4,g\Htr}
\nn
\\
&&
+\left(G_\delta/\bar G_\delta\right)^2\left(G_\Theta/\bar G_\Theta\right)^2
\bar{\cal T}_{5,g\Htr}
+\left(G_\delta/\bar G_\delta\right)\left(G_\Theta/\bar G_\Theta\right)^3
\bar{\cal T}_{6,g\Htr}
\nn
\\
&&
+\left(G_\Theta/\bar G_\Theta\right)^4
\bar{\cal T}_{7,g\Htr}, 
\ea
and,
\ba
\label{eq:estimatedFn_gH}
&&F_{g\Htr}(k,\mu) = \sum_{n=1}^{3} {\cal F}_{n,g\Htr}   \nn \\
&&= \left(G_\delta/\bar G_\delta\right)^2\left(G_\Theta/\bar G_\Theta\right)^2
\bar{\cal F}_{1,g\Htr}+\left(G_\delta/\bar G_\delta\right)\left(G_\Theta/\bar G_\Theta\right)^3\bar{\cal F}_{2,g\Htr} \nn \\
&&+\left(G_\Theta/\bar G_\Theta\right)^4
\bar{\cal F}_{3,g\Htr}. 
\ea
Here the expansion of the fiducial-cosmology-based correction terms $\bar{\cal A}_{n,g\Htr}$, $\bar{\cal B}_{n,g\Htr}$, $\bar{\cal T}_{n,g\Htr}$, and $\bar{\cal F}_{n,g\Htr}$ in terms of the matter power spectrum is shown in Appendix \ref{appendix:hybrid_cross}.

\subsection{Fisher analysis}
\label{subsec:fisher}

In this section, we demonstrate the constraining power of a joint galaxy and HI survey analysis based on our \emph{nonlinear} power spectrum expansion. For this purpose, we adopt the Fisher matrix formalism.
On the scales of our interest, the density and velocity fields are small enough to guarantee that the power spectrum is well described by our hybrid RSD model. It is well known that such an approach enables breaking the degeneracy between the growth and the bias for a single-tracer survey such as the galaxy survey \citep{2021PhRvD.104d3528S}. On the contrary, a power spectrum analysis based on the linear perturbation theory cannot break this degeneracy. In this section, we seek whether our non--linear perturbation theory for the two tracers $g$ and HI will enhance such a degeneracy-breaking capability and help constraining other parameters.

The actual isotropic power spectrum of the galaxy overdensity in real space, $\tilde{P}_{\delta_{g}\delta_{g}}(k)$, is affected by the discreteness of the tracer. Thus $\tilde{P}_{\delta_{g}\delta_{g}}(k)$ is given as \citep{McDonald_bias1,Saito2014,Hector_bias},
\bea
\tilde{P}_{\delta_{g}\delta_{g}}(k)&=&P_{\delta_g\delta_g}(k)+P_{\epsilon_g\epsilon_g},
\label{eq:pgg}
\eea
where $P_{\delta_{g}\delta_{g}}(k)$ and $P_{\epsilon_g\epsilon_g}$ are the theoretical real-space power spectrum and the stochastic contribution from the shot noise. We estimate $P_{\epsilon_g\epsilon_g}$ as a constant, given by the number density of the galaxy as a shot noise.
Similarly, the HI survey also has
\bea
\tilde{P}_{\delta_{\rm H}\delta_{\rm H}}(k)&=&P_{\delta_{\Htr}\delta_{\Htr}}(k)+P_{\epsilon_{\Htr}\epsilon_{\Htr}},
\label{eq:pHH}
\eea
where the shot noise term $P_{\epsilon_{\Htr}\epsilon_{\Htr}}$ arises from the discreteness of HI line emitters, and set as $P_{\epsilon_{\Htr}\epsilon_{\Htr}}=100\,h^{-3}\,{\rm Mpc}^{3}$. This value is based on the Schechter function fitted to the observed hydrogen mass function \citep{Zwaan2005}. Both $P_{\epsilon_g\epsilon_g}$ and $P_{\epsilon_{\Htr}\epsilon_{\Htr}}$ will set the minimum amount of uncertainties when assuming an ideal observation and included in the covariance matrix through Eqs.~(\ref{eq:covariance}--\ref{eq:noises}).

We are interested in analyzing galaxy survey and 21-cm mapping from the same sky, which makes the significant degeneracy among observables, $P_{gg}(k,\mu)$, $P_{g\Htr}(k,\mu)$ and $P_{\Htr\Htr}(k,\mu)$. The full covariance matrix for cross--correlation statistics is given by \citep{White:2008jy},
\begin{widetext}
\ba
\label{eq:cov_cross}
C_{\tilde{P}_{i}\tilde{P}_j}=\frac{1}{N_P}
\left(
	\begin{array}{ccc}
	2P_{gg}^2N_g^2      &    2P_{gg}P_{g\Htr}N_g     &      2P_{g\Htr}^2   \\
	2P_{gg}P_{g\Htr}N_g    &   \quad P_{g\Htr}^2+P_{gg}P_{\Htr\Htr}N_gN_{\Htr}  & \quad     2P_{g\Htr}P_{\Htr\Htr} N_{\Htr}  \\
	2P_{g\Htr}^2    &    2P_{g\Htr}P_{\Htr\Htr}N_{\Htr}     &      2P_{\Htr\Htr}^2N_{\Htr}^2   \\
	\end{array}
\right)
\label{eq:covariance}
\ea
\end{widetext}
where $i$ and $j$ denote $gg$, $g\Htr$ and $\Htr\Htr$ pairs, and $N_g$ and $N_{\Htr}$ are given by
\ba
N_g&=&\left(1+\frac{P^{(N)}_{g}}{P_{gg}}\right) , \nonumber \\
N_{\Htr}&=&\left(1+\frac{P^{(N)}_{\Htr}}{P_{\Htr\Htr}/p_S^2}\right) ,
\label{eq:noises}
\ea
where $P^{(N)}_{a}$ is the noise power spectrum of tracer $a$ inclusive of the shot, foreground and telescope noises and we assume
\ba
P^{(N)}_{a}&\simeq & P_{\epsilon_g \epsilon_g}=\frac{1}{\bar{n}_g} , \nonumber \\
P^{(N)}_{\Htr} &\simeq& P_{\epsilon_{\Htr} \epsilon_{\Htr}}+f_{\rm r}P_{\rm fore}+P_{\rm ant} \nonumber \\
&=& g_r \frac{1}{\bar{n}_{\Htr}} ,
\label{eq:noisepower}
\ea
where $\bar{n}_a$ is the comoving number density of tracer $a$, $P_{\rm fore}$ is the power spectrum of the foreground in the HI-survey frequency range, and $P_{\rm ant}$ is power spectrum due to the antenna temperature. We apply a fudge factor $f_{\rm r}$ ($\le 1$) to reflect an imperfect ``foreground removal'' process \citep{Liu2020} such that the resulting uncertainty would be due to the residual after removal. It is well known that $P_{\rm fore}$ is many orders of magnitude larger than the cosmic HI signal and thus a foreground removal is essential. Equivalently, we adopt an overall fudge factor $g_r$ ($\ge 1$) multiplied to $P_{\epsilon_g \epsilon_g}=1/\bar{n}_{\Htr} =100\,h^{-3}\,\rm Mpc^3$ to represent an ideal observation ($g_r =1$) to practical ones ($g_r\simeq 10$--$100$).

Let us adopt the SKA-MID antenna configuration to demonstrate the noise estimation. If one were to estimate the antenna noise, the radiometer equation can be used. We model the antenna noise as a Gaussian random noise with a standard deviation $\sigma_{\rm ant}$ in temperature given as
\begin{equation}
    \sigma_\mathrm{ant}=T_\mathrm{sys}\sqrt{\frac{4\pi f_\mathrm{sky}}{\Omega_{\rm beam}N_\mathrm{dish}t_\mathrm{obs}\Delta\nu}},
\label{eq:system}
\end{equation}
where $T_\mathrm{sys} \simeq 10$\,K is the system temperature, $\Omega_{\rm beam} \simeq 1.133\,\theta_{\rm FWHM}^2 \simeq 1.3\times10^{-3} \,\rm sr$ is the solid angle of the primary beam from 15-m dishes observing at $\sim 700\,\rm MHz$, the number of dishes $N_\mathrm{dish} = 133$ (representing the SKA-MID array), and $\Delta\nu = 1$\,MHz. For observing time $t_\mathrm{obs} = 200$\,hrs and the sky coverage $f_\mathrm{sky} \simeq 0.014$ (corresponding to a $1000^2\,\mathrm{Mpc}^2h^{-2}$ survey area at $z=1.03$), the noise level reaches $\sigma_\mathrm{ant} = 0.014$\,mK. Eq.~(\ref{eq:system}) is the estimated noise when a drift-scan, non-interferometric intensity mapping is performed. One can translate $\sigma_{\rm ant}$ into $P_{\rm ant}$ through \citep{2010ApJ...721..164S}.
\be
P_{\rm ant} = \frac{(k_B \sigma_{\rm ant} \Delta\nu)^2}{p^2_S}V_R,
\label{eq:Pant}
\ee
where
\ba
V_R &\simeq& 28800\,h^{-3}\,{\rm Mpc}^3 \nonumber \\
&&\left(\frac{R(z)}{2280\,h^{-1}\,{\rm Mpc}}\right)^2 \left(\frac{\theta_{\rm FWHM}}{0.0343}\right)^2 \left(\frac{\Delta \nu}{\rm MHz}\right)
\label{eq:VR}
\ea
is the ``pixel'' volume probed by the primary beam with $\theta_{\rm FWHM}(z)$ and the bandwidth $\Delta \nu$ when the line-of-sight comoving distance at $z$ is $R(z)$. SKA-MID with above antenna configuration then gives $P_{\rm ant}\simeq 60 \,h^{-3}\,{\rm Mpc^3}$, comparable to the expected shot noise $P_{\epsilon_{\Htr} \epsilon_{\Htr}}$.

The residual foreground, or the aftermath of removing the foreground dominated by the Galactic synchrotron radiation, is approximated as a white power spectrum:
\begin{equation}
    P_\mathrm{res} = f_\mathrm{r} \left(\frac{\bar{T}_\mathrm{sky}}{\hat{T}_{\rm sig}}\right)^2 V_R,
\label{eq:foreground}
\end{equation}
where $\bar{T}_\mathrm{sky} \simeq 10$\,K at the observing frequency of $\sim 700\,\rm MHz$, giving $\bar{T}_\mathrm{sky}/\hat{T}_{\rm sig}\sim 3\times10^4$. This white-noise form is to mimic the lack of correlation between the residual foreground and the galaxy survey. Efficient foreground removal will correspond to a very small $f_{\rm r}$, but we make a conservative assumption that the foreground removal is not perfect and the corresponding $g_{\rm r}$ in Eq.~(\ref{eq:noisepower}) can be as high as $\sim 100$.

The weight of $N_P$ in Eq.~(\ref{eq:cov_cross}) is given by,
\ba
N_P=\frac{V_{\rm survey}}{2(2\pi)^2} k^2\Delta k\Delta\mu\, .
\ea
The Fisher matrix is evaluated with 
\ba\label{eq:fisher_diag}
F_{\alpha\beta}= \sum_{k,\mu}\sum_{i,j}
\frac{\partial P_i(k,\mu)}{\partial \alpha}\Big[C_{\tilde{P}_i\tilde{P}_j}\Big]^{-1}\frac{\partial P_j(k,\mu)}{\partial \beta}. 
\ea
The full result is presented in the following section with three different cases: (1) galaxy auto--correlation, (2) HI auto--correlation, (3) the joint analysis of the galaxy auto--correlation, HI auto--correlation and galaxy--HI cross--correlation.


The parameter set is \{$\alpha_\perp$, $\alpha_{||}$, $f\sigma_8$, $\sigma_p$, $b_{g1}$, $b_{g2}$, $b_{\Htr 1}$, $b_{\Htr 2}$, $\sigma_8$, $A_S$, $x_{\rm HI}$\} where $\sigma_8$ and $f\sigma_8$ are linked to growth functions $G_\delta$ and $G_\Theta$, which are essential in our hybrid RSD formalism, via the following equations:
\ba
\sigma_8^2(z)&=&\frac{G_\delta^2(z)}{2\pi^2}\int W_8^2(k) k^2 P^i_{\delta\delta}(k) dk\,,  \\
(f\sigma_8)^2(z)&=&\frac{G_\Theta^2(z)}{2\pi^2}\int W_8^2(k) k^2 P^i_{\delta\delta}(k) dk\,. 
\ea
Here, the window function $W_8$ is given by
\ba
W_8=\frac{3j_1(kR_8)}{kR_8}\,, 
\ea
$j_1$ is the first-order spherical Bessel function, and $R_8=8
h^{-1}{\rm Mpc}$. All these parameters affect the power spectrum, and
we use the following fiducial values for our Fisher forecast: AP
parameters $\alpha_\perp$ and $\alpha_{||}$ are both unities, growth
factors are $\sigma_8=0.82$ and $f\sigma_8=0.46$, and the primordial
scalar amplitude is $A_S=2.15\times 10^{-9}$. These parameters are
varied around the fiducial values to construct the Fisher matrix. We
fix the shape parameter set for the primordial scalar perturbation, or
\{$\omega_m\equiv\Omega_m h^2$, $\omega_b \equiv\Omega_b h^2$,
$n_S$\}, to the Planck best-fit values \{0.14, 0.022, 0.96\} \citep{PLANK2015}. Nuisance parameters take the following fiducial values and are also varied for marginalization: the fiducial FoG damping parameter is $\sigma_p =3.3\hompc$ as estimated from the linear theory prediction, the galaxy bias parameters are \{$b_{g1}$, $b_{g2}$\}= \{1.45, 2.48\}, the 21cm bias parameters are \{$b_{\Htr 1}$, $b_{\Htr 2}$\}= \{1.1, 0.1\} ($b_{\Htr 1}$ roughly consistent with Ly$\alpha$ absorption system observation and hydrodynamical simulations: see \citealt{Villaescusa-Navarro_2018}, \citealt{Chen2019}, and \citealt{Castorina2019}; $b_{\Htr 2}$ is poorly constrained and thus we adopt an arbitrary non-zero value), and $x_{\rm HI}=0.027$.  In summary, there are 5 cosmological parameters, 5 nuisance parameters and 1 astrophysical parameter which are varied without any prior information.

\section{Results: constraints on cosmological parameters}
\label{sec:result}

\begin{figure*}
\begin{center}
\resizebox{3.in}{!}{\includegraphics{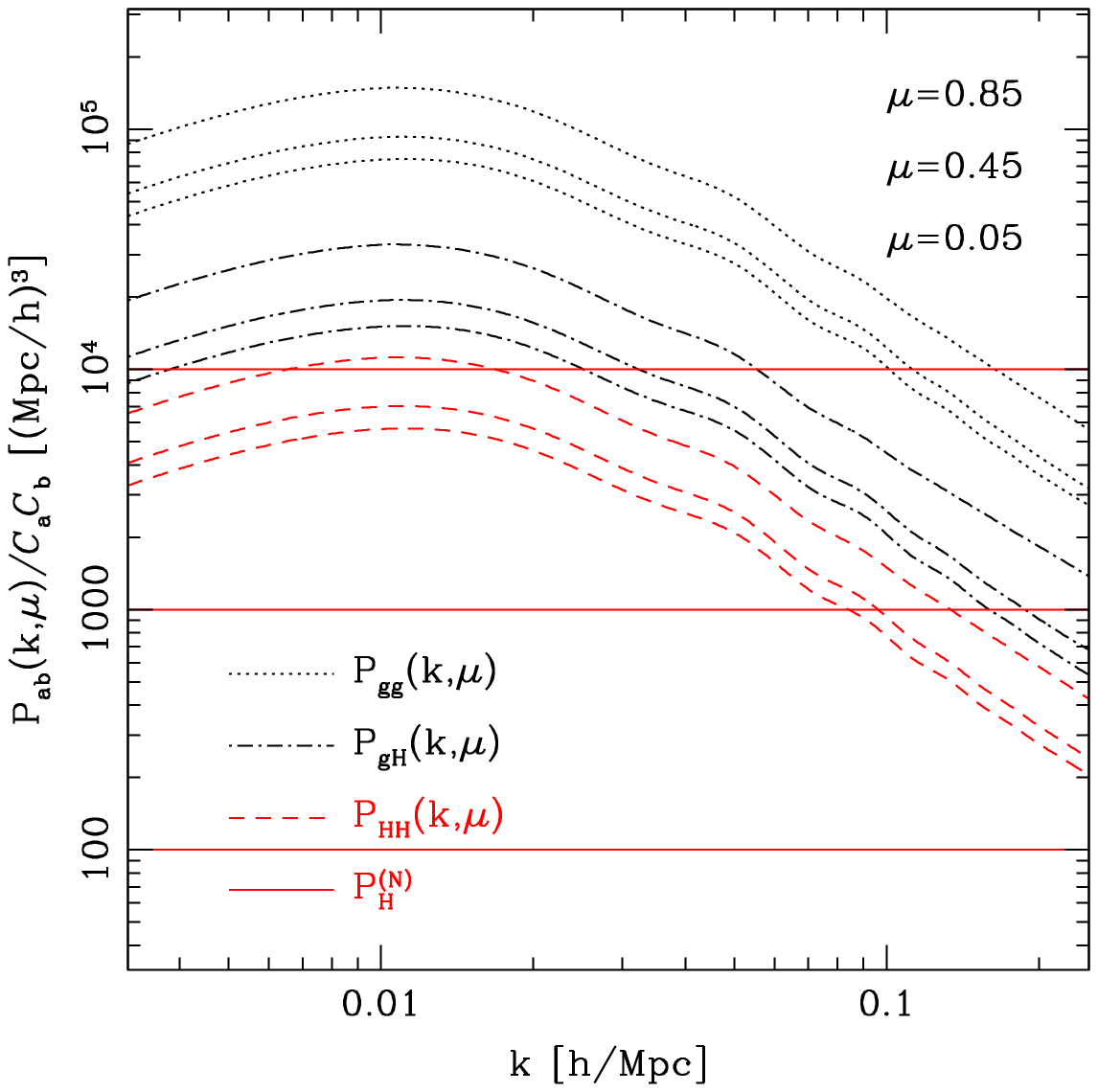}}\hfill
\resizebox{3.in}{!}{\includegraphics{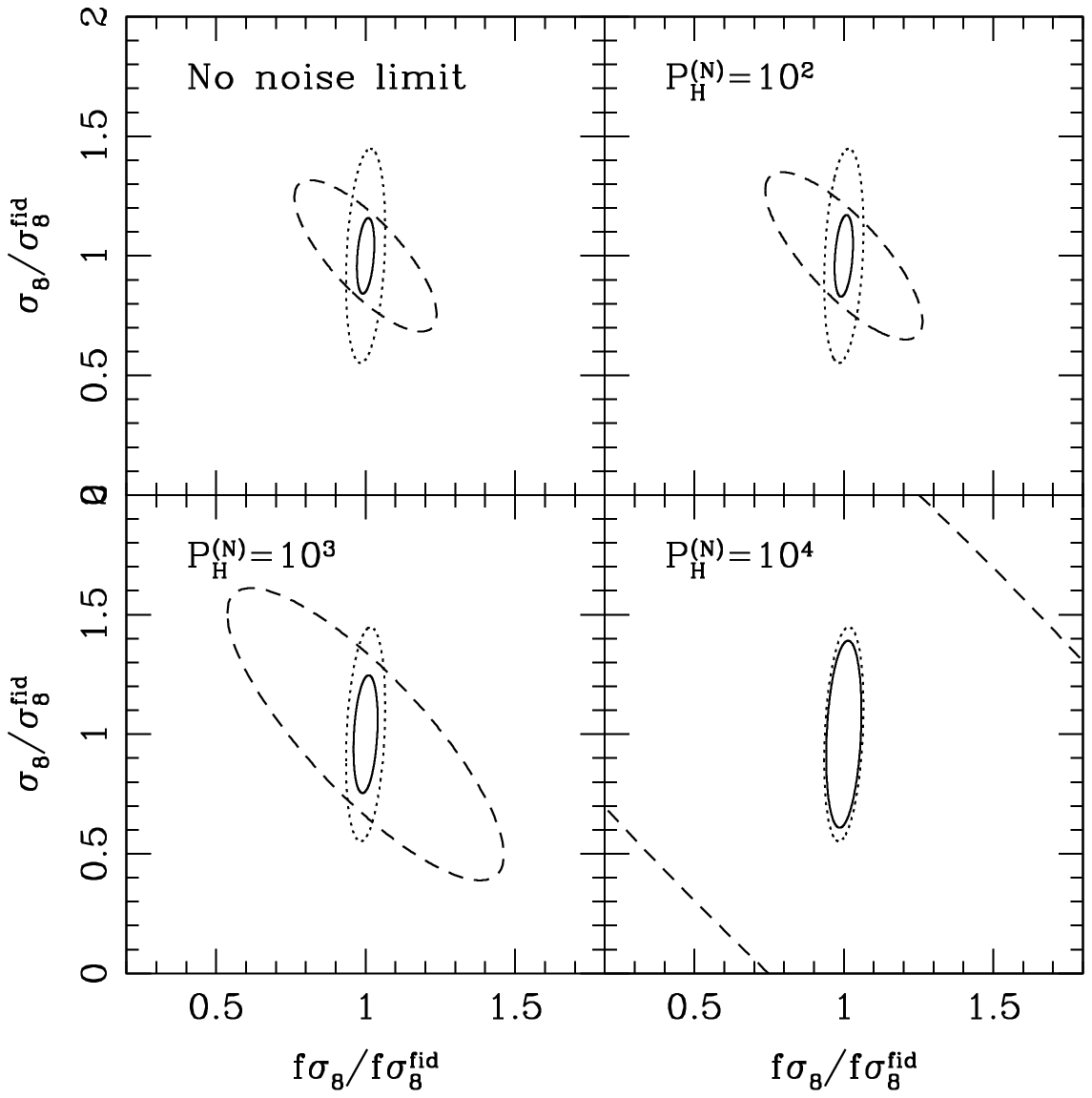}} 
\vspace*{0.5cm}
\end{center}
\caption{{\it Left}: Toy model power spectra (each $P_{ab}$ normalized by $\mathscr{C}_a \mathscr{C}_b$): $P_{gg}$ (dotted), $P_{g\Htr}$ (dot-dashed) and $P_{\Htr\Htr}$ (dashed) together with varying HI noise-power spectra (red, solid) $P^{(N)}_{\Htr}=(10^2,10^3,10^4) (\mpcoh)^3$. The anisotropic power spectra are presented by varying $\mu=(0.05,0.45,0.85)$ from bottom to top. {\it Right}: The contour plots between $\sigma_8$ and $f\sigma_8$ are presented with galaxy auto spectrum case, HI auto spectrum case and full cross spectra case presented by dotted, dashed, and solid contours. The given foreground noise level along with no foreground noise limit is denoted in each panel.}
\label{fig:fs8s8}
\end{figure*}

\subsection{Precise simultaneous probe of $\sigma_8$ and $f\sigma_8$}
\label{sec:sigma8}

The large-scale redshift-space power spectrum can be precisely modeled into the mildly non-linear regime.
While the density growth function, parameterized by $\sigma_8$, and the linear bias $b_{a1}$ ($a=\{g,\,{\rm H}\}$) are highly degenerate on linear scales, they become mildly separable near the Baryon Acoustic Oscillation (BAO) scale.
As density fluctuations cross the quasi-linear threshold, modes in varying scales undergo non-linear coupling, imprinting distinct non-Gaussian signatures.
Because this non-linear evolution in redshift space is governed by both density and velocity growth parameters, it yields distinguishable observational effects.

Consequently, $b_{a1}$ and $\sigma_8$ can, in principle, be measured separately using RSD. However, residual degeneracies between the coherent bias and $\sigma_8$ can still degrade constraints on other cosmological parameters.
Although increasing $k_{\rm max}$ past $0.2 \, h\,{\rm Mpc}^{-1}$ might further alleviate this degeneracy, it is valuable instead to evaluate the impact of full $\sigma_8$ marginalization on $f\sigma_8$ within the existing scale cutoff, thereby highlighting the intrinsic synergy of the combined galaxy and 21-cm intensity mapping analysis.

The left panel of Figure~\ref{fig:fs8s8} presents auto- and cross-correlation power spectra together with varying foreground-noise power spectrum $P_{\rm H}^{(N)} \in \{10^2, 10^3, 10^4\} \mpcohthree$, covering optimistic to conservative scenarios.
For context, existing MeerKAT studies report $P_{\rm H}^{(N)} \gtrsim 3000 \mpcohthree$ at $k \sim 0.1 \hompc$ using $N_{\rm fg}=4$ principal components for foreground subtraction \citep{MeerKLASS2023}.\footnote{While \citet{MeerKLASS2023} extend $N_{\rm fg}$ to 30, power loss becomes significant even at $N_{\rm fg}=10$ for $k \sim 0.1\hompc$. We adopt $N_{\rm fg}=4$ to provide a practical estimate of foreground contamination while preserving large-scale power for cross-correlation.}
The galaxy auto-power spectra $P_{gg}(k,\mu)$ (with $b_{g1}=1.5$) are shown as black dotted curves.
The 21-cm auto-power spectra $P_{\Htr\Htr}(k,\mu)$ (with $b_{H1}=1.1$) are plotted as red dashed curves, and the cross-spectra $P_{g\Htr}(k,\mu)$ as black dot-dashed curves.

The right panel of Figure~\ref{fig:fs8s8} presents the expected $1\sigma$ cosmological constraints in the $\sigma_8$--$f\sigma_8$ plane deduced from surveys at $1.0 < z < 1.2$, illustrating the potential of the power-spectrum analysis under varying foreground noise levels. Because $\sigma_8$ is probed via higher-order terms in the perturbative expansion, its constraint is inherently weaker than parameters probed primarily at leading order; nevertheless, a single-tracer spectroscopic survey can simultaneously capture density and velocity information to constrain $\sigma_8$ within a few percent. For this analysis, leading- and next-to-leading-order local galaxy biases are varied independently alongside $\sigma_8$, while non-local bias is fixed by the updated $b_{a1}$. Although these biases are poorly constrained, their uncertainties do not significantly propagate into other cosmological parameters except for $\sigma_8$.

This figure also highlights the degeneracies associated with the structure growth parameter for auto-correlation analysis from each tracer. $\sigma_8$ is highly correlated with the linear growth rate $f\sigma_8$, leading to a strong parameter degeneracy when using the power spectrum alone (dotted and dashed contours for galaxy and HI surverys, respectively). This degeneracy is driven by the conservative cutoff $k_{\rm max} = 0.2 \, h\,{\rm Mpc}^{-1}$, below which the damping term varies monotonically along the line-of-sight direction $\mu$ and mimics variations in $f$. Due to the strong $\sigma_8$--$b_1$ degeneracy, $f\sigma_8$ cannot be constrained beyond the $10\%$ level using the galaxy  auto-correlation alone. A similar and even worse limitation applies to the HI auto-correlation, where the constraint depends heavily on the efficacy of the foreground removal pipeline; the sensitivity approaches the galaxy-galaxy case only when the 21-cm signal is cleanly recovered down to non-linear scales.

However, the joint analysis (solid contours) substantially improve the constraint on $\sigma_8$, reducing the uncertainty to below $5\%$.
Furthermore, the velocity growth rate $f\sigma_8$ is now constrained to sub-percent precision, a result that remains remarkably robust against uncertainties in the foreground cleaning procedure. 
The underlying mechanism for such an improvement lies in scale-dependent non-linearities and the synergy of dual tracers. As noted earlier, the linear bias $b_{X1}$ mixes with $\sigma_8$ through higher-order corrections in the quasi-linear regime, contrasting sharply with the full degeneracy between the two parameters on the fully linear scale.
Consequently, the response of the power spectrum to variations in $\sigma_8$ becomes dependent on the specific value of $b_{x1}$.
Moreover, employing different tracers with distinct non-linear responses rotates the covariance between $\sigma_8$ and $b_{X1}$ (Fig.~\ref{fig:fs8s8}), breaking the degeneracy even further and allowing for a precise determination of $f\sigma_8$.

For context, ELGs trace relatively massive halos ($\sim 10^{12} M_\odot$) with an expected bias of $b_g \approx 1.4$, while 21-cm HI emission originates from much lower-mass halos ($\sim 10^{10} M_\odot$) with an assumed bias of $b_{\rm HI} = 1.1$.
Because higher-order perturbative terms scale differently with these distinct bias values, their auto-correlations yield misaligned covariance ellipses between $\sigma_8$ and $f\sigma_8$.
Cross-correlating them successfully separates the two parameters.

\subsection{Benefit for distance measurements}
\label{sec:distance}

Precise measurement of the BAO scale is essential for determining the geometric distances, $D_A(z)$ and $H^{-1}(z)$, via the AP effect. While the acoustic structure imprinted on the power spectrum inherently depends on the underlying cosmology, its primordial counterpart is tightly constrained by CMB anisotropies. By anchoring our analysis with WMAP and Planck priors, the BAO serves as a robust standard ruler. In the linear regime ($k\lesssim 0.2\hompc$), the galaxy power spectrum clearly preserves this acoustic signature. Furthermore, although galaxy bias introduces uncertainty into the clustering amplitude, it leaves the characteristic BAO scales largely unaffected.

\begin{figure}
\begin{center}
\resizebox{3.in}{!}{\includegraphics{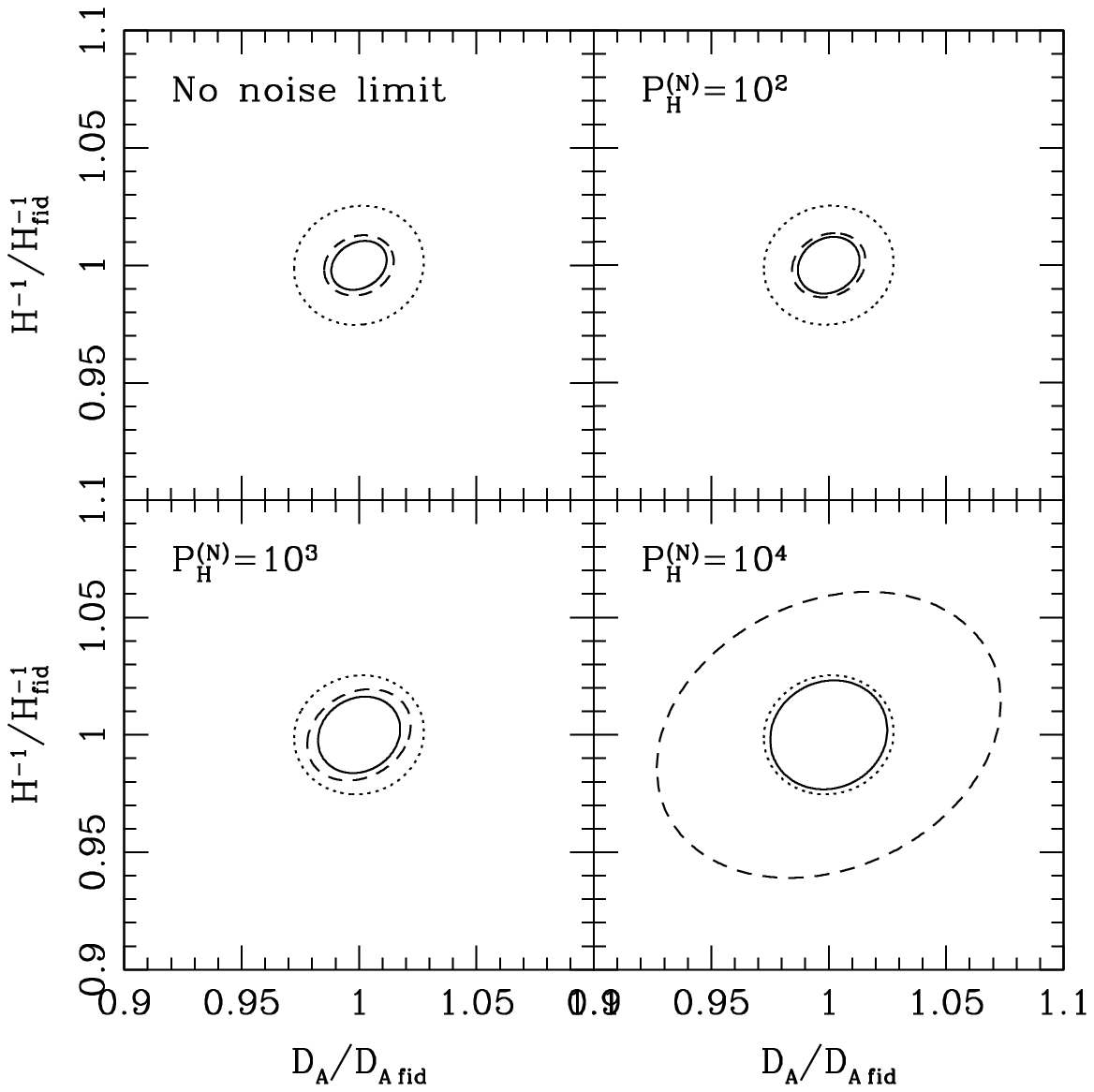}}
\vspace*{0.5cm}
\end{center}
\caption{Contour plots for $D_{A}/D_{A,\rm fid}$ and $H^{-1}/H^{-1}_{\rm fid}$, comparing constraints derived from the galaxy auto power spectrum (dotted contours), the HI auto power spectrum (dashed contours), and the full joint analysis (solid contours). The assumed HI noise power is indicated in each panel.}
\label{fig:distance}
\end{figure}

Figure~\ref{fig:distance} (x-axis) illustrates the forecasted statistical errors for the angular diameter distance $D_A$. For a DESI-like survey, the constraint on $D_A$ from the galaxy power spectrum (dotted contours) alone is comparable to that from an ideal 21-cm experiment (dashed contours). However, the constraining power of the 21-cm experiment rapidly degrades as the foreground noise power increases. It is worth noting that if parameters such as the growth rate ($f\sigma_8$) and linear galaxy bias were known a priori, $D_A$ could be measured with sub-percent accuracy. In practice, these parameters are instead highly degenerate with $D_A$, particularly in the transverse limit ($\mu\rightarrow 0$). 

Because the 21-cm signal-to-noise ratio in the zero-noise limit (minimum $P_{\rm H}^{(N)}$) exceeds that of the galaxy power spectrum, the 21-cm data alone dominates the cosmological constraints (as seen in Figure~\ref{fig:distance}). Consequently, combining the galaxy survey with the 21-cm experiment (solid contours) offers negligible improvement over the constraints based on the HI power spectrum alone. This saturation of constraining power aligns with previous forecasts. For instance, \citet{10.1093/mnras/staa2914} found comparable results using angular cross-correlation power spectra. Our analysis yields a constraining power of roughly 2--3 per cent for $P_{\rm H}^{(N)}\simeq 10^{4}\mpcohthree$. We note that as foreground removal improves, the precision increases more slowly than it does for the HI auto-correlation case. Furthermore, while \citet{10.1093/mnras/staa2914} focused solely on the cross-correlation---reporting a 2.7 per cent mean error across 10 redshift bins---our prediction comprehensively incorporates HI auto-correlations, galaxy auto-correlations, and their cross-correlation with a single redshift bin. The overall joint analysis can provide about a factor of two improvement in constraining $D_A$.

Constraining the radial distance $H^{-1}$  with the joint analysis shows similar improvement seen in the case of $D_A$. One can also observe the same trend: such an improvement is mostly due to the accuracy from the prevailing auto-power spectrum. Overall, the precision of the full joint analysis for $D_A$ and $H^{-1}$ is driven by whichever auto-power spectrum --- galaxy or HI --- yields the higher precision.

\subsection{Astrophysical implication: neutral hydrogen fraction}
\label{sec:xHI}

\begin{deluxetable}{cccc}
\tablewidth{0pt}
\tablecaption{Error forecast on $x_{\rm HI}$}
\tablehead{
  \colhead{$P_{\rm H}^{({\rm N})}$} & \colhead{galaxy-galaxy} & \colhead{21-cm-21-cm} & \colhead{galaxy-21-cm}}
\startdata
  no noise & ----- & 13$\%$ & 0.32$\%$ \\
  $10^2 \mpcoh^3$ & ----- & 15$\%$ & 0.34$\%$ \\
  $10^3 \mpcoh^3$ & ----- & 28$\%$ & 0.57$\%$ \\
  $10^4 \mpcoh^3$ & ----- & 140$\%$ & 2.6$\%$ \\
\enddata
\label{tab:sigmaHI}
\tablecomments{The fractional errors $\sigma(x_{\rm HI})/x_{\rm HI}$ are presented with varying toy model foreground noises ($P_{\rm H}^{({\rm N})}$).}
\end{deluxetable}
Our result shows that the joint analysis on power spectra (gg, gH and HH) can very strongly constrain $x_{\rm HI}$, as is shown in Table~\ref{tab:sigmaHI}. Percent level or even the sub-percent level we expect in $x_{\rm HI}$ estimation far surpasses the currently available estimations giving $\sigma(x_{\rm HI})/x_{\rm HI}\sim 1$. Depending on the degree of the systematic uncertainties ($P_{\rm H}^{({\rm N})}$, not restricted to the sky brightness but also inclusive of the antenna systematics and shot noises due to the finite number of galaxies and 21-cm sources), the 21-cm auto-correlation is found to constrain $x_{\rm HI}$ to a few tens of per cent when $P_{\rm H}^{({\rm N})}\lesssim 10^{3}\mpcohthree$. However, the joint analysis (cross+auto) is found to constrain $x_{\rm HI}$ to sub-percent levels when $P_{\rm H}^{({\rm N})}\lesssim 10^{3}\mpcohthree$, and to a few-percent level even when $P_{\rm H}^{({\rm N})}\simeq 10^{4}\mpcohthree$. We show our sanity check on the smallness of our estimated $\sigma(x_{\rm HI})$ in Appendix~\ref{appendix:Fab}.
This is a very useful result because $x_{\rm HI}$ encapsulates the post-reionization astrophysics as a whole which is still poorly constrained.

\begin{figure*}
\gridline{
\fig{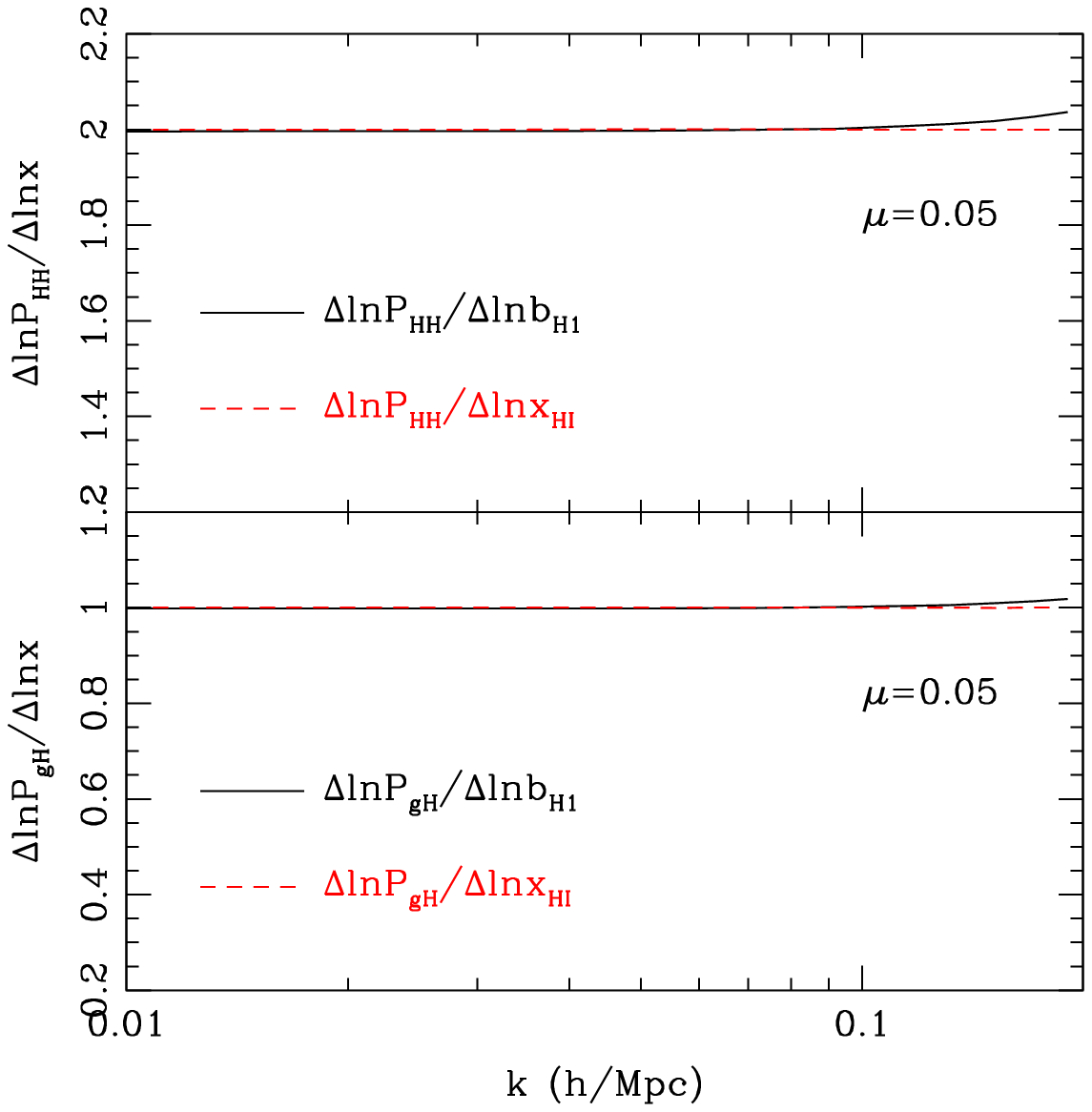}{0.45\textwidth}{(a)}
\fig{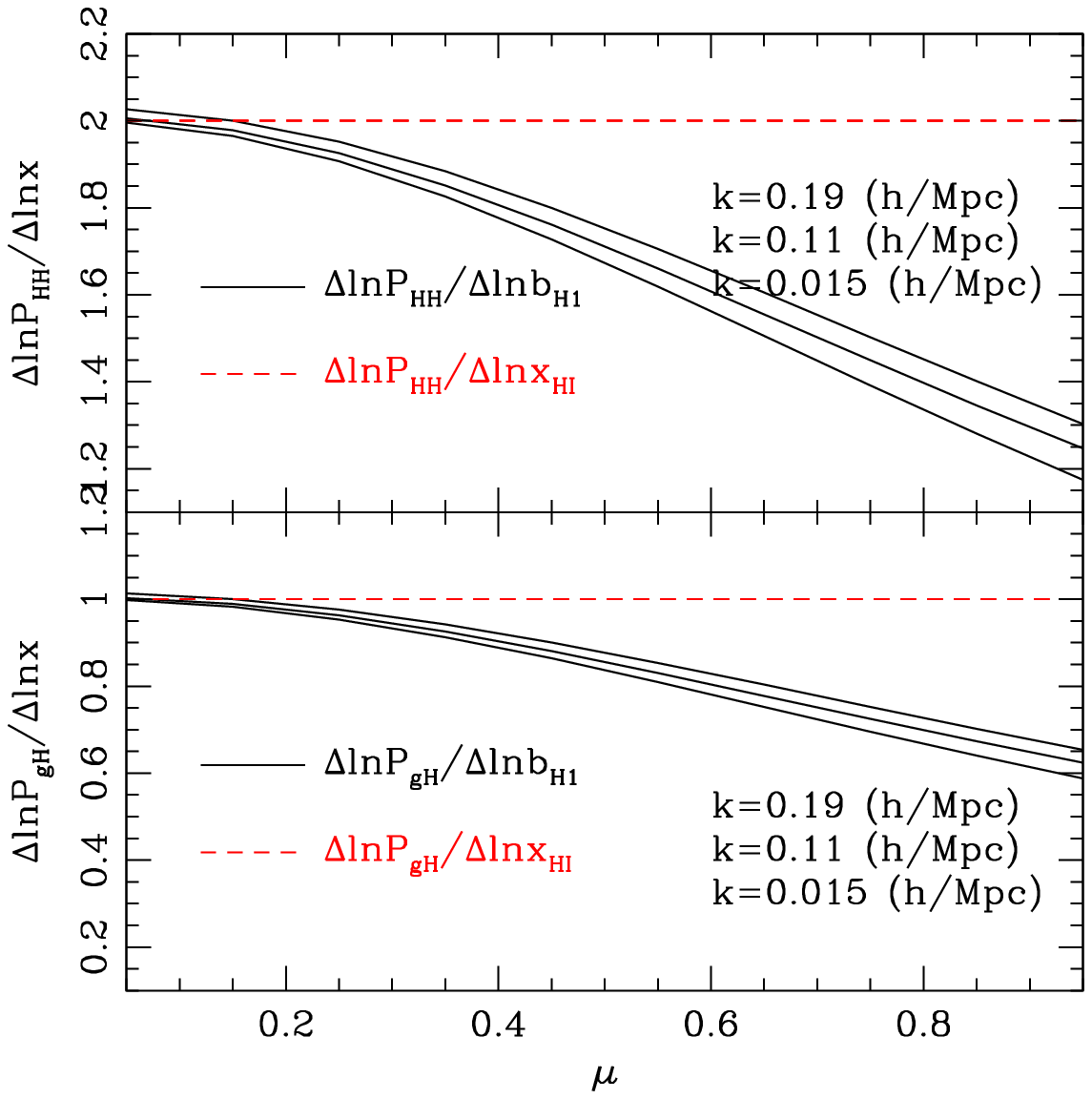}{0.45\textwidth}{(b)}
}
\caption{(a) Variations of power spectra of HI-HI (top) and g-HI (bottom) against the HI bias $b_{\rm HI}$ and the HI fraction $x_{\rm HI}$ in terms of $k$ at $\mu=0.05$. (b) Same as (a) but also with varying $k$ (smallest to largest from top to bottom curves), now in terms of $\mu$. Overall, dependence of the anisotropic power spectra on $\mu$ and $b_{\rm HI}$ makes it possible for our nonlinear prescription to break the degeneracy between $b_{\rm HI}$ and $x_{\rm HI}$.}
\label{fig:dpdx}
\end{figure*}

We believe that such a high accuracy in $x_{\rm HI}$ estimation is thanks to the capability to break the degeneracy between $b_{\rm HI}$ (more accurately, the leading order bias $b_{{\rm H}{1}}$) and $x_{\rm HI}$. We demonstrate this by examining the dependence of the anisotropic power spectra on $b_{\rm HI}$ and $\mu$. As seen in Figure~\ref{fig:dpdx}, while the impact of $x_{\rm HI}$ on power spectra is trivial (working as a simple multiplier to power spectra), the impact of $b_{\rm HI}$ is significant. Dependence of the power spectra on $b_{\rm HI}$ gets amplified as $\mu$ becomes larger. Therefore, according to our nonlinear formalism, observing the anisotropic power spectra can constrain $b_{\rm HI}$. This then breaks the usual degeneracy between $b_{\rm HI}$ and $x_{\rm HI}$ that has been reported on previous studies (e.g. \citealt{MeerKLASS2023}). 


Such a tight constraint on $x_{\rm HI}$ could be one of the most accurate, and can help us understand how post-reionization physics has proceeded. So far, constraints come mainly from observing the Lyman-$\alpha$ forest \citep{Rao2006,Grasha2020} with values ranging from $\Omega_{\rm HI}(z)\equiv x_{\rm HI}(z) \Omega_{\rm H,0} = (0.7\pm 0.3)\times 10^{-3}$ \citep{Rao2006} to $\Omega_{\rm HI}(z)= (1.2\pm 0.3)\times 10^{-3}$ \citep{Grasha2020} at $z\simeq 1$. Direct detection of 21-cm emission objects can only give lower limits to $x_{\rm HI}$ at the moment due to the limited sensitivity of telescopes compared to the smallness of their usual flux \citep{ALFALFA018,Chowdhury2020,FASHI2024}. In other words, currently $\sigma(x_{\rm HI})/x_{\rm HI}\simeq 1$ and we expect the measure of $P_{g\rm H}$ for the joint analysis with $P_{gg}$ and $P_{\rm HH}$ can improve the constraint to a very ideal level.

One could in principle further constrain the physics of the hydrogen clumps with such a tight constraint on $x_{\rm HI}$. Suppose that we could model the hydrogen clump distribution with a simple hydrogen mass function $dn/dM_{\rm HI}$ (comoving clump number density per unit hydrogen clump mass), possibly in a form such as the Schechter function, and assume the rest of the universe fully ionized, then we will have
\beq
x_{\rm HI}=\frac{1}{\Omega_{\rm H,0}\rho_{\rm crit,0}}\int_{M_{\rm HI,Min}} M_{\rm HI}\frac{dn}{dM_{\rm HI}}dM_{\rm HI},
\eeq
where $M_{\rm HI,Min}$ is the minimum hydrogen clump mass.
Therefore, once the mass function is modelled to a certain accuracy and extrapolated to the small-mass end that may not be detected as individual galaxies both in the galaxy survey and the 21-cm survey, we may be able to further constrain the minimum hydrogen clump mass $M_{\rm HI,Min}$. $M_{\rm HI,Min}$ is expected to be determined primarily by the Jeans-mass filtering of the heated intergalactic medium by photo-heating, and thus this can be extended to the understanding of the post-reionization process.


\section{Conclusion}
\label{sec:conclusion}


Although cross-correlation only modestly improves constraints on cosmic distances---roughly doubling the accuracy even in the most ideal scenario---it dramatically improves measurements of the density and velocity growth functions ($\sigma_8$ and $f\sigma_8$). In this work, we demonstrate that these growth parameters can be precisely measured through a full joint analysis that combines the auto- and cross-correlations of a galaxy survey and an HI line-intensity mapping (LIM) survey.

If we could observe the underlying dark matter density field directly without bias, both growth parameters could be extracted purely from the redshift-space anisotropy. In this idealized case, the transverse limit ($\mu \rightarrow 0$) is unaffected by redshift-space distortions (RSD), allowing us to measure $\sigma_8$ directly as a baseline amplitude. Once this baseline is established, $f\sigma_8$ can be determined by quantifying the RSD squashing effect as $\mu$ increases toward the line of sight.

In reality, however, dark matter fluctuations are probed indirectly using biased tracers, such as galaxies or HI emission. Because tracer bias is perfectly degenerate with the clustering amplitude, the baseline measurement at $\mu \rightarrow 0$ yields the combined parameter $b\sigma_8$ rather than $\sigma_8$ alone. Consequently, a single auto-correlation measurement cannot simultaneously break this degeneracy to isolate both $\sigma_8$ and $f\sigma_8$, necessitating the use of cross-correlation statistics.

The measured power spectrum is highly sensitive to the variation of overall amplitude, and the signal to noise has been improved with recent wide deep field experiment. If all other cosmological and nuisance parameters are pre--determined, then $x_{\rm HI}$ will be determined in the sub--percentage level. However, the constraint on a parameter with an featureless overall amplitude contribution becomes much poorer in marginalization with other parameters. For instance, $x_{\rm HI}$ parameter is most degenerate with galaxy bias $b_{\rm HI}$.


The anisotropy-driven squeezing effect induced by RSD provides a unique opportunity to break the degeneracy between $x_{\rm HI}$ and $b_{\rm HI}$. In traditional galaxy surveys, the peculiar velocity field is independent of galaxy bias. It is true also for the HI survey. However, the mean temperature $\hat{T}_{\rm sig}$, which is proportional to $x_{\rm HI}$, scales both the density and velocity perturbations equally. As a result, variations in $x_{\rm HI}$ exert a featureless amplitude shift on the power spectrum that persists equally at any angle ($\mu$). Because $x_{\rm HI}$ and $b_{\rm HI}$ possess mutually different angular dependencies in redshift space, this RSD anisotropy effectively breaks the degeneracy between the two parameters.

Nevertheless, such a degeneracy breaking will not be possible with the HI auto power spectrum alone. There is an additional degeneracy between $\sigma_8$ and $b_{\rm HI}$ which is broken by the cross--correlation. In summary, the chain of degeneracy breaking is as follows: 1) galaxy bias is separable from $\sigma_8$ by cross--correlation statistics through distinct non--linearity, 2) $x_{\rm HI}$ parameter becomes distinguished from galaxy bias through anisotropy in redshift distortion.

A major advantage of the joint analysis of galaxy and HI surveys is the highly accurate estimation of $x_{\rm HI}$, which we demonstrated  can be achieved at the sub-percent level. Consequently we might be able to probe the temporal evolution of $x_{\rm HI}$ very accurately. We used a rather restricted range of redshift, $1.0<z<1.2$, in this study. If one were to extend the analysis, roughly put, $\sigma(x_{\rm HI})\propto \Delta r^{-0.5}$, where $\Delta r$ is the longitudinal comoving distance in a given redshift bin, assuming a simple sample variance. Therefore, about a percent-level accuracy is warranted for $x_{\rm HI}(z)$ measurement at e.g. $\sim 20$ redshift bins from $2\gtrsim z \gtrsim 0$. Further study is warranted.

\section*{Acknowledgments}

We would like to thank Takahiro Nishimichi for useful discussions and comments. Numerical calculations were performed using a high performance computing cluster in the Korea Astronomy and Space Science Institute. F.S acknowledges the support from the National SKA Program of China (2022SKA0110200 and 2022SKA0110202). KA is supported by the National Research Foundation of Korea (NRF) RS-2021-NR058956, RS-2025-16302968 and the Korea Astronomy and Space Science Institute under the R\&D program (Project No. 2025-9-844-00) supervised by the Korea AeroSpace Administration. YSS is supported by the National
Research Foundation of Korea (NRF) grant funded by
the Korea govbernment (MIST) RS-2021-NR058702.

\appendix

\section{Description of leading--order hybrid RSD model for matter power spectrum}
\label{appendix:hybrid_DM_RSD}
This section guides the reader through the computation of each term in the square brackets of Eq.~(\ref{eq:Pkred_final}) utilizing the hybrid approach to model the RSD effect. While comprehensive details are available in \citet{Song_2018} and \citet{2021PhRvD.104d3528S}, we briefly outline the application for a few selected terms to ensure that this paper remains self-contained. Appendix \ref{appendix:hybrid_DM_RSD} details the first four terms within the brackets of Eq.~(\ref{eq:Pkred_final}), while Appendix \ref{appendix:hybrid_cross} addresses the remaining higher--order terms: $A_{ab}(k,\mu)$, $B_{ab}(k,\mu)$, $T_{ab}(k,\mu)$, and $F_{ab}(k,\mu)$. Consistent with our earlier notation, barred quantities indicate the fiducial cosmology, whereas unbarred quantities represent the new cosmology.

First, major contributions to the terms $P_{\delta_a\delta_a}$, $P_{\Theta\Theta}$ and $P_{\delta_a\Theta}$ in Eq.~(\ref{eq:Pkred_final}) originate from $P_{XY}(k,z)$ $(X,Y=\delta\,\,\mbox{or}\,\,\Theta)$. For instance, $P_{\delta\delta}$ contributes to $P_{\delta_g\delta_g}$ in Eq.~(\ref{eq:pgg_bias}),  and $P_{\delta\Theta}$ contributes to $P_{\delta_g\delta_\Theta}$ in Eq.~(\ref{eq:pgt_bias}).
As introduced in Section~\ref{sec:method}, we adopt a hybrid RSD modeling approach, combining RegPT-based theoretical templates $\bar P^{\rm th}_{XY}(k,z)$ with simulation measurements $\bar P^{\rm mea}_{XY}(k,z)$. The theoretical templates can be expressed in terms of multi-point propagators up to the two--loop order as follows:
\ba
&&\bar P^{\rm th}_{XY}(k,z) = \bar \Gamma_{X, \rm reg}^{(1)}(k,z)\bar \Gamma_{Y, \rm reg}^{(1)}(k,z)\bar P^{i}(k)+2\int\frac{d^3\vec q}{(2\pi)^3}\bar \Gamma_{X, \rm reg}^{(2)}(\vec q, \vec k-\vec q,z) \bar \Gamma_{Y, \rm reg}^{(2)}(\vec q, \vec k-\vec q,z) \bar P^i(q) \bar P^{i}(|\vec k-\vec q|) 
\nn
\\
&&\quad
+6\int\frac{d^3\vec pd^3\vec q}{(2\pi)^6} \bar \Gamma_{X, \rm reg}^{(3)}(\vec p,\vec q, \vec k-\vec p-\vec q,z)\bar \Gamma_{Y, \rm reg}^{(3)}(\vec p,\vec q, \vec k-\vec p-\vec q,z) \bar P^i(p) \bar P^i(q) \bar P^i(|\vec k-\vec p-\vec q|).
\label{eq:pk_RegPT}
\ea
Here, $\bar P^i$ is the initial linear power spectrum calculated by the publicly available Boltzmann solver \texttt{CAMB} \citep{Lewis:1999bs}, and $\bar \Gamma_{X, \rm reg}^{(m)}$ is the $(m+1)$-point propagator with the following expression:
\begin{equation}
\bar \Gamma_{X, \rm reg}^{(m)} = {\rm exp}\left(-\bar G_{\delta}^2\,\bar \gamma \right) \bar G_X \, \sum_n \, \bar G_\delta^{n-1} \, \bar{\cal C}^{(m)}_{n, X}(\bar \gamma),
\end{equation}
which is a comprehensive explicit form of  Eqs.~(24-26) of \citet{RegPT} and can be found in  Section 2.2 of \citet{Song_2018} as well. Note that theoretical predictions of $\bar P^{\rm th}_{XY}$ up to the two-loop order are achieved using the publicly available code {\tt RegPT} \citep{2014ascl.soft04012T} with the Planck best-fit parameter values \citep{PLANK2015} as described in Section~\ref{subsec:fisher}.
To overcome the limitations of the perturbative approach at low redshifts for small scales $k > 0.1\,h\,{\rm Mpc}^{-1}$ (especially at $z\,=\,0.5$ as reported in \citet{Song_2018}), we characterize the difference between the measured and predicted power spectra as a residual $\bar P_{XY}^{\rm res}=\bar P^{\rm mea}_{XY}(k,z) - \bar P^{\rm th}_{XY}(k,z)$. We then transform this residual from the fiducial to the new cosmology using a growth--function dependence $G_XG_YG_\delta^4$:
\ba
P^{\rm mea}_{XY}(k,z)=P^{\rm th}_{XY}(k,z)+\frac{G_XG_YG_\delta^4}{\bar G_X \bar G_Y \bar G_\delta^4}\bar P^{\rm res}_{XY}(k,z).
\label{eq:pk_hybrid_new}
\ea

Second, the remaining terms in Eq.~(\ref{eq:pgg_bias}) and Eq.~(\ref{eq:pgt_bias}) correspond to one-loop corrections due to higher order bias terms and their explicit expressions are provided in Appendix B of \citet{Hector_bias}. For example, for the sake of completeness, the second term $P_{b2,\delta}$ in Eq.~(\ref{eq:pgg_bias})  can be obtained from the one-loop integrals\footnote{Here, "one-loop integral" means the integral over one internal momentum.} as follows:
\ba
P_{b2,\delta} = \int \frac{d^3\bfq}{(2\pi)^3}P_{\rm m}^{\rm L}(q)P_{\rm m}^{\rm L}(|\bfk-\bfq|)\mathscr{F}^{\rm SPT}_{2}(\bfq, \bfk-\bfq)\, ,
\label{eq:P_hibias_ex}
\ea
where the $\mathscr{F}^{\rm SPT}_{2}$ kernel is given by
\ba
\mathscr{F}^{\rm SPT}_{2}(\bfk_i, \bfk_j) = \frac{5}{7}+\frac{1}{2}\frac{\bfk_i\cdot\bfk_j}{k_i k_j}\Bigl(\frac{k_i}{k_j}+\frac{k_j}{k_i}\Bigr) + \frac{2}{7}\Bigl[\frac{\bfk_i\cdot\bfk_j}{k_i k_j}\Bigr]^2.
\ea
As above, we transform the theoretical prediction at the fiducial cosmology into a new cosmology using the following scaling relation $P_{\rm m}^{\rm L} = G^2_{\delta}/\bar G^2_{\delta} \, \bar P_{\rm m}^{\rm L}$. We apply the same procedure to the second term $P_{b2,\Theta}$ in Eq.~(\ref{eq:pgt_bias}) using a  different kernel $\mathscr{G}^{\rm SPT}_{2}$ which is given by
\ba
\mathscr{G}^{\rm SPT}_{2}(\bfk_i, \bfk_j) = \frac{3}{7}+\frac{1}{2}\frac{\bfk_i\cdot\bfk_j}{k_i k_j}\Bigl(\frac{k_i}{k_j}+\frac{k_j}{k_i}\Bigr) + \frac{4}{7}\Bigl[\frac{\bfk_i\cdot\bfk_j}{k_i k_j}\Bigr]^2.
\ea

\section{Description of higher--order hybrid RSD model for auto- and cross- power spectrum for two tracers}
\label{appendix:hybrid_cross}
Next, we present a method for estimating the higher--order clustering terms of the tracers. This is achieved by generalizing the dark matter auto--power spectrum expressions--detailed in Sec. 2.3 of \citet{Song_2018}--to the cross--power spectrum between tracer $a$ and tracer $b$. These terms are then incorporated into Eq. (\ref{eq:Pkred_final}) to compute the complete cross--power spectrum. We begin our derivation by considering the term $A_{ab}$. 
\ba
A_{ab}(k,\mu) && = j_1\int d^3\bfx e^{i\bfk\cdot\bfx}\langle A_1A_2^aA_3^b\rangle_c \nonumber\\
&& = j_1 \int d^3\bfx e^{i\bfk\cdot\bfx}\langle (u_z-u_z')(b_{a1}\delta+\nabla_z u_z)(b_{b1}\delta'+\nabla_z u'_z)\rangle_c,
\label{eq:higherorder_A}
\ea
where $A_1, A^{a}_2$ and $A^{b}_3$ are defined in Section~\ref{subsec:hybrid_DM_RSD_cross}. Based on its explicit form, the term $A_{ab}$ can be decomposed into four distinct components. Below, we provide these expressions evaluated within the fiducial cosmological model: 
\be
  \bar A_{ab}(k,\mu) =  \sum_{n=1}^{4} \bar {\cal A}_{n,ab}.
\label{eq:A_term_tracer}
\ee
Here, barred quantities denote values evaluated within the fiducial cosmological model. The explicit form of $\bar {\cal A}_n$ is given by:
\ba
\bar {\cal A}_{1,ab}&=& 2j_1\,\int d^3\bfx \,\,e^{i\bfk\cdot\bfx}\,\,\langle u_z(\bfr) \delta(\bfr) \delta(\bfr')\rangle_c b_{a1} b_{b1} , \nn
\label{eq:A_1}\\
\bar {\cal A}_{2,ab}  &=& j_1\,\int d^3\bfx \,\,e^{i\bfk\cdot\bfx}\,\,\langle u_z(\bfr)\delta(\bfr)\,\nabla_zu_z(\bfr')\rangle_c (b_{a1}+b_{b1}), \nn
\label{eq:A_2}\\
\bar {\cal A}_{3,ab}  &=& j_1\,\int d^3\bfx \,\,e^{i\bfk\cdot\bfx}\,\,\langle u_z(\bfr)\,\nabla_zu_z(\bfr)\delta(\bfr')\rangle_c (b_{a1}+b_{b1}), \nn
\label{eq:A_3}\\
\bar {\cal A}_{4,ab} &=& 2j_1\,\int d^3\bfx \,\,e^{i\bfk\cdot\bfx}\,\,\langle u_z(\bfr)\,\nabla_zu_z(\bfr)\,\nabla_zu_z(\bfr')\rangle_c,
\label{eq:Anab}
\ea
where $b_{a1}$ and $b_{b1}$ are linear biases for tracers $a$ and $b$, with the subscripts $a$ and $b$ denoting either $g$ or $\rm H$. Following the methodology of \citet{Zheng_2016}, we extract these terms from $N$-body simulations. To extrapolate these measurements to other cosmological models, we adopt a scaling ansatz analogous to the one used for the power spectrum $P_{\rm XY}$. Specifically, we assume that the scale dependence of each term is largely insensitive to the underlying cosmology, allowing us to evaluate these terms simply by rescaling the simulated data. We postulate that the time evolution is primarily governed by the leading-order growth factors of $u_z$ and $\delta$. Consequently, the term $A_{ab}$ for an arbitrary cosmological model can be expressed as:
\ba
A_{ab}(k,\mu) = \sum_{n=1}^{4} {\cal A}_{n,ab} &&=\left(G_\delta/\bar G_\delta\right)^2\left(G_\Theta/\bar G_\Theta\right) \bar{\cal A}_{1,ab} \nn \\
&& +\left(G_\delta/\bar G_\delta\right)\left(G_\Theta/\bar G_\Theta\right)^2
\bar{\cal A}_{2,ab}+\left(G_\delta/\bar G_\delta\right)\left(G_\Theta/\bar G_\Theta\right)^2 \bar{\cal A}_{3,ab} \nn \\
&&+\left(G_\Theta/\bar G_\Theta\right)^3
\bar{\cal A}_{4,ab}.
\label{eq:estimatedAn_ab}
\ea
We apply the same procedure to other higher--order corrections. Starting with the explicit form of the term $B_{ab}$, we have 
\bea
B_{ab}(k,\mu) && = j_1^2\,\int d^3\bfx \,\,e^{i\bfk\cdot\bfx} \langle A_1A_2^a\rangle_c\,\langle A_1A_3^b\rangle_c,\nonumber\\
&& = j^2_1 \int d^3\bfx e^{i\bfk\cdot\bfx} \langle(u_z-u_z')(b_{a1}\delta+\nabla_z u_z)\rangle_c \langle(u_z-u_z')(b_{b1}\delta'+\nabla_z u'_z)\rangle_c. \nn
\eea
We decompose this correction in the fiducial cosmology into several components. Applying the scaling ansatz yields the following predictions:
\begin{eqnarray}
  \bar B_{ab}(k,\mu) =  \sum_{n=1}^{3} \bar {\cal B}_{n,ab}, \nn
\end{eqnarray}
where the explicit forms of $\bar {\cal B}_{n,ab}$ are given below:
\ba
\bar {\cal B}_{1,ab}&=& - j_1^2\,\int d^3\bfx \,\,e^{i\bfk\cdot\bfx}\,\,\langle u_z(\bfr') \delta(\bfr)\rangle_c \langle u_z(\bfr)\delta(\bfr')\rangle_c b_{a1} b_{b1} , \nn
\label{eq:B_1}\\
\bar {\cal B}_{2,ab}  &=& - j_1^2\,\int d^3\bfx \,\,e^{i\bfk\cdot\bfx}\,\,\langle u_z(\bfr')\delta(\bfr) \rangle_c \langle u_z(\bfr)\nabla_zu_z(\bfr')\rangle_c (b_{a1}+b_{b1}), \nn
\label{eq:B_2}\\
\bar {\cal B}_{3,ab} &=& - j_1^2\,\int d^3\bfx \,\,e^{i\bfk\cdot\bfx}\,\,\langle u_z(\bfr')\,\nabla_zu_z(\bfr) \rangle_c \langle u_z(\bfr)\nabla_zu_z(\bfr')\rangle_c.
\label{eq:Bnab}
\ea
Thus, the $B_{ab}$ term in a general cosmological model is expressed as
\ba
\label{eq:estimatedBn_ab}
B_{ab}(k,\mu) =\sum_{n=1}^{3} {\cal B}_{n,ab} &&= \left(G_\delta/\bar G_\delta\right)^2\left(G_\Theta/\bar G_\Theta\right)^2
\bar{\cal B}_{1,ab}
\nn
\\
&&+\left(G_\delta/\bar G_\delta\right)\left(G_\Theta/\bar G_\Theta\right)^3
\bar{\cal B}_{2,ab}+\left(G_\Theta/\bar G_\Theta\right)^4
\bar{\cal B}_{3,ab}.
\ea
The explicit expression for $T_{ab}(k,\mu)$ is
\bea
T_{ab}(k,\mu) && = \frac{1}{2} j_1^2\,\int d^3\bfx \,\,e^{i\bfk\cdot\bfx}\,\,\langle A_1^2A_2^aA_3^b\rangle_c \nonumber \\
&& = \frac{1}{2} j_1^2 \int d^3\bfx \,\,e^{i\bfk\cdot\bfx}\,\,\langle (u_z-u'_z)^2 (b_{a1}\delta+\nabla_zu_z)(b_{b1}\delta'+\nabla_zu'_z)\rangle_c. \nn
\eea
This equation can also be decomposed into seven components in the fiducial cosmology:
\begin{eqnarray}
  \bar T_{ab}(k,\mu) = \sum_{n=1}^{7} \bar {\cal T}_{n,ab}, \nn
\end{eqnarray}
where the explicit forms of $\bar {\cal T}_{n,ab}$ are given below:
\ba
\bar {\cal T}_{1,ab}&=&   j_1^2\,\int d^3\bfx \,\,e^{i\bfk\cdot\bfx}\,\,\langle u_z(\bfr)u_z(\bfr)\delta(\bfr)\delta(\bfr') \rangle_c b_{a1} b_{b1},  \nn \\
\bar {\cal T}_{2,ab}  &=& j_1^2\,\int d^3\bfx \,\,e^{i\bfk\cdot\bfx}\,\,\langle u_z(\bfr)u_z(\bfr)\delta(\bfr)\,\nabla_zu_z(\bfr')  \rangle_c (b_{a1}+b_{b1}),  \nn \\
\bar {\cal T}_{3,ab}  &=&  j_1^2\,\int d^3\bfx \,\,e^{i\bfk\cdot\bfx}\,\,\langle u_z(\bfr)u_z(\bfr)\,\nabla_zu_z(\bfr)\delta(\bfr')  \rangle_c (b_{a1}+b_{b1}),  \nn \\
\bar {\cal T}_{4,ab} &=& j_1^2\,\int d^3\bfx \,\,e^{i\bfk\cdot\bfx}\,\,\langle u_z(\bfr)u_z(\bfr)\,\nabla_zu_z(\bfr)\,\nabla_zu_z(\bfr')  \rangle_c,  \nn \\
\bar {\cal T}_{5,ab}&=&  - j_1^2\,\int d^3\bfx \,\,e^{i\bfk\cdot\bfx}\,\,\langle u_z(\bfr')u_z(\bfr)\delta(\bfr)\delta(\bfr') \rangle_c b_{a1} b_{b1},  \nn \\
\bar {\cal T}_{6,ab}  &=& -2 j_1^2\int d^3\bfx \,\,e^{i\bfk\cdot\bfx}\,\,\langle u_z(\bfr')u_z(\bfr)\delta(\bfr)\,\nabla_zu_z(\bfr')  \rangle_c (b_{a1}+b_{b1}),  \nn \\
\bar {\cal T}_{7,ab} &=& - j_1^2\,\int d^3\bfx \,\,e^{i\bfk\cdot\bfx}\,\,\langle u_z(\bfr')u_z(\bfr)\,\nabla_zu_z(\bfr)\,\nabla_zu_z(\bfr')  \rangle_c.
\label{eq:Tnab}
\ea
These components are then rescaled with respect to the barred quantity in the fiducial cosmology using the growth functions, which leads to the following prediction in a general cosmology:
\ba
\label{eq:estimatedTn_ab}
T_{ab}(k,\mu)= \sum_{n=1}^{7} {\cal T}_{n,ab} &&= \left(G_\delta/\bar G_\delta\right)^2\left(G_\Theta/\bar G_\Theta\right)^2
\bar{\cal T}_{1,ab} \nn \\
&&+\left(G_\delta/\bar G_\delta\right)\left(G_\Theta/\bar G_\Theta\right)^3
\bar{\cal T}_{2,ab}+\left(G_\delta/\bar G_\delta\right)\left(G_\Theta/\bar G_\Theta\right)^3
\bar{\cal T}_{3,ab} \nn \\
&&+\left(G_\Theta/\bar G_\Theta\right)^4
\bar{\cal T}_{4,ab}+\left(G_\delta/\bar G_\delta\right)^2\left(G_\Theta/\bar G_\Theta\right)^2
\bar{\cal T}_{5,ab} \nn \\
&&+\left(G_\delta/\bar G_\delta\right)\left(G_\Theta/\bar G_\Theta\right)^3
\bar{\cal T}_{6,ab}+\left(G_\Theta/\bar G_\Theta\right)^4
\bar{\cal T}_{7,ab}. 
\ea
Lastly, the explicit expression for $F_{ab}(k,\mu)$ is given by:
\bea
F_{ab}(k,\mu) && = -j_1^2\,\int d^3\bfx \,\,e^{i\bfk\cdot\bfx}\,\,\langle u_z u_z'\rangle_c\langle A_2^aA_3^b\rangle_c \nonumber \\
&& = -j_1^2\int d^3\bfx \,\,e^{i\bfk\cdot\bfx}\,\,\langle u_z u'_z\rangle_c \langle (b_{a1}\delta+\nabla_zu_z)(b_{b1}\delta'+\nabla_zu'_z)\rangle_c \nn
\eea
This can be decomposed in the fiducial cosmology into three components:
 \begin{eqnarray}
  \bar F_{ab}(k,\mu) = \sum_{n=1}^{3} \bar {\cal F}_{n,ab}, \nn
\end{eqnarray}
where the explicit forms of $\bar {\cal F}_{n,ab}$ are given below:
\ba
\bar{\cal F}_{1,ab}&=& -j_1^2\,\int d^3\bfx \,\,e^{i\bfk\cdot\bfx}\,\,\langle u_z u_z'\rangle_c\langle \delta(\bfr)\delta(\bfr')\rangle_c b_{a1} b_{b1},  \nn\\
\bar{\cal F}_{2,ab}&=& -j_1^2\,\int d^3\bfx \,\,e^{i\bfk\cdot\bfx}\,\,\langle u_z u_z'\rangle_c\langle \delta(\bfr)\,\nabla_zu_z(\bfr')\rangle_c (b_{a1}+b_{b1}),  \nn \\
\bar{\cal F}_{3,ab}&=& -j_1^2\,\int d^3\bfx \,\,e^{i\bfk\cdot\bfx}\,\,\langle u_z u_z'\rangle_c\langle \,\nabla_zu_z(\bfr)\,\nabla_zu_z(\bfr')\rangle_c.
\label{eq:Fnab}
\ea
Thus, the term $F_{ab}$ in a general cosmological model is expressed as:
\ba
\label{eq:estimatedFn_ab}
F_{ab}(k,\mu) = \sum_{n=1}^{3} {\cal F}_{n,ab} &&=\left(G_\delta/\bar G_\delta\right)^2\left(G_\Theta/\bar G_\Theta\right)^2
\bar{\cal F}_{1,ab}\nn\\
&&+\left(G_\delta/\bar G_\delta\right)\left(G_\Theta/\bar G_\Theta\right)^3
\bar{\cal F}_{2,ab}+\left(G_\Theta/\bar G_\Theta\right)^4
\bar{\cal F}_{3,ab}.
\ea
Again, the quantities $\bar{\cal B}_{n,ab}$, $\bar{\cal F}_{n,ab}$, and  $\bar{\cal T}_{n,ab}$ are measured in the fiducial cosmological model. Both $\sigma_8$ and $f\sigma_8$ are derived from the measured growth functions of $G_\delta$ and $G_\Theta$.

\section{An analytical estimation of the diagonal Fisher matrix element for HI}
\label{appendix:Fab}
To better understand the marginalized posteriors of the parameters, we calculate the analytical form of the Fisher matrix $F_{\alpha \beta}$:
\ba
F_{\alpha \beta}=\sum_{k,\mu}\mathcal{F}_{\alpha \beta}(k,\mu),
\ea
where
\ba
\mathcal{F}_{\alpha \beta}(k,\mu)\equiv\sum_{i,j}\frac{\partial P_i(k,\mu)}{\partial \alpha}\Big[C_{\tilde{P}_i\tilde{P}_j}\Big]^{-1}\frac{\partial P_j(k,\mu)}{\partial \beta}.
\ea
Expanding this expression, we find:
\ba
\mathcal{F}_{\alpha\beta}(k,\mu) &&=
\frac{N_p}{2(P_{g\Htr}^2-P_{gg}P_{\Htr\Htr}N_{g}N_{\Htr})^2} \left\{\frac{\partial P_{gg}}{\partial \alpha}\frac{\partial P_{gg}}{\partial \beta}P_{\Htr\Htr}^2 N_{\Htr}^2\right. -2\frac{\partial P_{gg}}{\partial \alpha}\frac{\partial P_{g\Htr}}{\partial \beta}P_{g\Htr}P_{\Htr\Htr}N_{\Htr} +\frac{\partial P_{gg}}{\partial \alpha}\frac{\partial P_{\Htr\Htr}}{\partial \beta}P_{g\Htr}^2 \nonumber \\
&&-2\frac{\partial P_{g\Htr}}{\partial \alpha}\frac{\partial P_{gg}}{\partial \beta}P_{g\Htr}P_{\Htr\Htr}N_{\Htr} +2\frac{\partial P_{g\Htr}}{\partial \alpha}\frac{\partial P_{g\Htr}}{\partial \beta}\left(P_{g\Htr}^{2}+P_{gg}P_{\Htr\Htr}N_{g}N_{\Htr}\right) -2\frac{\partial P_{g\Htr}}{\partial \alpha}\frac{\partial P_{\Htr\Htr}}{\partial \beta}P_{g\Htr}P_{gg}N_{g}  \nonumber \\
&&+\frac{\partial P_{\Htr\Htr}}{\partial \alpha}\frac{\partial P_{gg}}{\partial \beta}P_{g\Htr}^2 -2\frac{\partial P_{\Htr\Htr}}{\partial \alpha}\frac{\partial P_{g\Htr}}{\partial \beta}P_{gg}P_{g\Htr}N_{g} +\left.\frac{\partial P_{\Htr\Htr}}{\partial \alpha}\frac{\partial P_{\Htr\Htr}}{\partial \beta} P_{gg}^{2}N_{g}^{2}\right\},
\label{eq:fisheranal}
\ea
where we have omitted the dependence of the power spectra on $k$ and $\mu$ for brevity.

The diagonal elements are relevant to our marginalized posterior estimations, giving a good sanity check through the Cram\'{e}r-Rao inequality:
\ba
\sigma(\alpha)\ge \frac{1}{\sqrt{F_{\alpha \alpha}}}.
\ea
In particular, we have checked the surprisingly high accuracy in the estimated $x_{\rm HI}$ by
directly calculating, using Eq.~(\ref{eq:fisheranal}),
\ba
\mathcal{F}_{x_{\rm HI}x_{\rm HI}}(k,\mu)&&=\frac{N_p}{2(P_{g\Htr}^2-P_{gg}P_{\Htr\Htr}N_{g}N_{\Htr})^2} \left\{ 2\left(\frac{\partial P_{g\Htr}}{\partial x_{\rm HI}}\right)^2 \left(P_{g\Htr}^{2}+P_{gg}P_{\Htr\Htr}N_{g}N_{\Htr}\right) \right. \nn \\
&& -4\frac{\partial P_{g\Htr}}{\partial x_{\rm HI}} \frac{\partial P_{\Htr\Htr}}{\partial x_{\rm HI}} P_{gg}P_{\Htr\Htr}N_{g} \left.+\left(\frac{\partial P_{\Htr\Htr}}{\partial x_{\rm HI}}\right)^{2}P_{gg}^{2}N_{g}^{2}\right\}.
\label{eq:FHIHI}
\ea
This sanity check confirms that the reported value for $\sigma(x_{\rm HI})$ in Section~\ref{sec:xHI} is indeed correct.

\bibliography{reference} 

\end{document}